\documentclass[12pt]{article}

\usepackage{amssymb}
\usepackage{amsmath}
\usepackage{amscd}
\usepackage{latexsym}

\usepackage{graphicx}
\usepackage{array}

\usepackage{cite}

\newcommand{\be}{\begin{equation}}
\newcommand{\ee}{\end{equation}}

\newcommand{\dlt}{\delta}

\newcommand{\bt}{\beta}
\newcommand{\vp}{\varphi}

\newcommand{\al}{\alpha}
\newcommand{\ra}{\rightarrow}

\newcommand{\lbd}{\lambda}

\newcommand{\cH}{{\cal H}}

\newcommand{\rgl}{\rangle}
\newcommand{\lgl}{\langle}

\begin{document}

\begin{center}
{\Large{\bf Information Processing by Networks of Quantum Decision Makers} \\ [5mm]

V.I.~Yukalov$^{1,2,*}$, E.P.~Yukalova$^{1,3}$, and D.~Sornette$^{1,4}$} \\ [3mm] 

{\it 
$^1$Department of Management, Technology and Economics, ETH Z\"urich, \\
Swiss Federal Institute of Technology, Z\"urich CH-8032, Switzerland \\ [2mm]

$^2$Bogolubov Laboratory of Theoretical Physics, \\
Joint Institute for Nuclear Research, Dubna 141980, Russia \\ [2mm]

$^3$Laboratory of Information Technologies, \\
Joint Institute for Nuclear Research, Dubna 141980, Russia \\ [2mm]

$^4$Swiss Finance Institute, c/o University of Geneva, \\
40 blvd. Du Pont d'Arve, CH 1211 Geneva 4, Switzerland  }

\end{center}

\vskip 1cm

\begin{abstract}
We suggest a model of a multi-agent society of decision makers taking decisions being 
based on two criteria, one is the utility of the prospects and the other is the attractiveness
of the considered prospects. The model is the generalization of quantum decision theory, 
developed earlier for single decision makers realizing one-step decisions, in two principal 
aspects. First, several decision makers are considered simultaneously, who interact with 
each other through information exchange. Second, a multistep procedure is treated, when 
the agents exchange information many times. Several decision makers exchanging
information and forming their judgement, using quantum rules,  form a kind of a quantum 
information network, where collective decisions develop in time as a result of information 
exchange. In addition to characterizing collective decisions that arise in human societies, such 
networks can describe dynamical processes occurring in artificial quantum intelligence 
composed of several parts or in a cluster of quantum computers. The practical usage of 
the theory is illustrated on the dynamic disjunction effect for which  three quantitative 
predictions are made: (i) the probabilistic behavior of decision makers at the initial stage 
of the process is described; (ii) the decrease of the difference between the initial
prospect probabilities and the related utility factors is proved; (iii) the existence of a 
common consensus after multiple exchange of information is predicted. The predicted 
numerical values are in very good agreement with empirical data. 
\end{abstract}

\vskip 1cm

{\parindent =0pt


\vskip 1cm
$^*${\bf Corresponding author}: V.I. Yukalov

\vskip 2mm

{\bf E-mail}: yukalov@theor.jinr.ru     }

\newpage

\section{Introduction and Related Literature}

Modeling the behavior of multi-agent social systems and social networks has recently 
attracted a substantial amount of activities of researchers, as can be inferred from the 
review articles \cite{Kiesling_81,Perc_112,Perc_113} and numerous original papers, 
of which we can cite just a few recent 
\cite{Zhou_114,Fabbri_115,Lee_116,Heiberger_117,Gao_118}. High interest to such 
a modeling is due to two reasons. First, being able of describing the behavior of societies 
is of great importance by itself. Second,  modeling the collective interactions of autonomous 
multientity systems has already been widely envisioned to be a powerful paradigm for 
multi-agent computing.

Nowadays, there exists an extensive literature on decision making in multi-agent systems, 
which can be categorized into three main directions: (i) Studies of how the agents, inhabiting 
a shared environment, can decide their actions through mutual negotiations. There can be 
two types of agents, cooperative and self-interested. Cooperative agents cooperate with 
each other to reach a common goal \cite{Moehlman_92}. Self-interested agents try to maximize 
their own payoff without concern to the global good, choosing the best negotiation strategy 
for themselves \cite{Parsons_94,Jennings_95}. (ii) Studies of how a network of agents with 
initially different opinions can reach a collective decision and take action in a distributed 
manner \cite{Gal_97}. (iii) Studies of how the dependence among multi-agents can lead to 
emerging social structures, such as groups and agent clusters \cite{Shoham_99}.  

The main goal of a decision process in a multi-agent system is to find the optimal policy
that maximizes expected utility or expected reward for either a single agent or for the 
society as a whole. Various models of multi-agent systems can be found in the books
\cite{Russel_100,Pearl_101,Weiss_102,Wooldridge_103,Shoham_104,Weiss_105}. 

The maximization of expected utility, expected reward, or other functionals is based 
on the assumption that agents are perfectly rational and no restrictions on computational
power and available resources are imposed. However, since Simon \cite{Simon_106},
it is well known that only {\it bounded rationality} can exist, so that any real decision maker,
in addition to having limited computational power and finite time for deliberations, is 
subject to such behavioral effects as irrational emotions, subconscious feelings, and 
subjective biases \cite{Simon_106,Simon_107,Rubinstein_108,Gigerenzer_109}.  

The behavioral effects become especially important when decisions are made under
uncertainty. There are two sides of uncertainty that can be caused either by objective 
lack of complete knowledge or by subjective preferences and biases. Even when the 
agents in the society are assumed to possess complete knowledge, they cannot 
become absolutely rational decision makers due to the inherent dual property of 
human behavior that combines conscious evaluation of utility with subconscious feelings 
and emotions \cite{Loomes_116}. 

Thus, real decision making is a complex procedure of dual nature, simultaneously integrating
the rational, conscious, objective evaluation of utility with behavioral effects, such as  
irrational emotions, subconscious feelings, and  subjective biases. This feature of realistic 
decision making can be called {\it rational-irrational duality}, or {\it conscious-subconscious duality},
or {\it objective-subjective duality}. Keeping in mind these points, we can call this feature
the {\it behavioral duality of decision making}.    

As a result of this behavioral duality, a correct description of decision making in a real 
multi-agent system has to deal with two sides -- maximization of expected reward, and taking 
account of behavioral effects \cite{Steunebrink_110,Balke_111,Bulling_112}. To take the latter
into account, several modifications of utility theory have been suggested, such as prospect 
theory, weighted-utility theory, regret theory, optimism-pessimism theory, dual-utility theory, 
ordinal-independence theory, and quadratic-probability theory, whose description can 
be found  in the review articles \cite{Camerer_113,Machina_4,Machina_5,Baillon_6}.
However, such so-called nonexpected utility models listed in reviews 
\cite{Machina_4,Machina_5,Baillon_6} have been refuted as being merely descriptive 
and having no predictive power \cite{Birnbaum_7,Birnbaum_8,Safra_9,Alnajjar_10,Alnajjar_11}. 
A more detailed discussion can be found in Refs. \cite{YS_1,YS_119,YS_2}.  

As has been shown by Safra and Segal \cite{Safra_9}, none of non-expected utility theories 
can resolve all problems and  paradoxes typical of decision making of humans. The best 
that could be achieved is a kind of fitting for interpreting just one or, in the best case, a few
problems, while the other remained unexplained. In addition, spoiling the structure of expected 
utility theory results in the appearance of several complications and inconsistences. As has 
been concluded in the detailed analysis of Al-Najjar and Weinstein \cite{Alnajjar_10,Alnajjar_11}, 
any variation of the classical expected utility theory ``ends up creating more paradoxes and 
inconsistences than it resolves''.

Stochastic decision theories are usually based on deterministic decision theories
complemented by random variables with given distributions \cite{Cockett_114,Blavatskyy_44}. 
Therefore, such stochastic theories inherit the same problems as deterministic theories 
embedded into them. Moreover, stochastic theories are descriptive, containing fitting 
parameters that need to be defined from empirical data. In addition, different stochastic 
specifications of the same deterministic core theory may generate very different, and 
sometimes contradictory, conclusions \cite{Loomes_115}. 

One more difficulty in modeling decision making of real humans is that they often vary
their decisions, under the same invariant expected utility, after information exchange 
between decision makers, as has been observed in many empirical studies 
\cite{Charness_66,Blinder_67,Cooper_68,Charness_69,
Charness_70,Chen_71,Charness_72,Kuhberger_73}. This implies that agent
interactions through information exchange can influence decision makers emotions,
without touching their evaluation of utility. 

In all previous works, behavioral effects, when being considered, have been treated as
stationary. The principal difference of the present paper from all previous publications
is that we develop a model that takes into account the dual nature of decision making, and
allows for the description of dynamical behavioral effects caused by agent interactions
through information exchange.

\section{Main Features of Quantum Approach}

To take into account the dual nature of decision making, in the previous papers 
\cite{YS_1,YS_119,YS_2}, we have formulated Quantum Decision Theory (QDT) as 
a mathematical approach for describing decision making under uncertainty. This approach 
generalizes classical utility theory \cite{Neumann_3} to the cases of decisions under 
strong uncertainty, when utility theory fails. The necessity of developing a new approach 
has been  justified by numerous empirical observations proving the failure of classical 
utility making in realistic situations.

The mathematical basis of the QDT is the generalization of the von Neumann
\cite{Neumann_12} theory of quantum measurements to the case of inconclusive quantum
measurements \cite{YS_13} involving composite events with intermediate operationally
untestable steps \cite{YS_14,YS_15}. In decision making, the intermediate inconclusive
events characterize subconscious feelings and deliberations of a decision maker. It 
is this dual feature of decision making, including conscious logical reasoning and
subconscious intuitive feelings, which explains the successful application of quantum
techniques to describing human decision making, without the necessity of assuming any 
quantum nature of decision makers. The mathematics of quantum theory turns out to be
well suited for describing the intrinsic conscious-subconscious duality of human 
cognition \cite{Helland_86,Khan_65}.

It is worth mentioning that, after Bohr \cite{Bohr_16} put forward the idea that the dual 
nature of consciousness requires the use of quantum description, a number of attempts 
have been made to apply quantum rules to cognition, as can be inferred from the review 
works \cite{YS_17,Khrennikov_18,Busemeyer_19,Haven_20,Bagarello_21,Agrawal_22,
Sornette_23,Ashtiani_24}. The previous attempts, however, were limited by models fitted 
to particular cases and having no general mathematical structure valid for arbitrary events. 
Moreover, as has been recently shown \cite{Boyer_25,Boyer_26}, these models
contradict empirical facts.  

Contrary to this, our QDT \cite{YS_1,YS_119,YS_2} is general, being formulated for arbitrary 
composite events. Its mathematics is based on the theory of quantum measurements
\cite{YS_13,YS_14,YS_15}, which allows it to be used also for the problem of 
creating artificial quantum intelligence \cite{YS_2,YS_27,YS_28}. As has been recently
demonstrated \cite{YS_1,YS_119,YS_2}, QDT is the sole theory that provides the 
possibility of making {\it quantitative} predictions, without fitting parameters, even in such 
difficult situations when classical decision making is not even applicable qualitatively.

In our previous papers \cite{YS_1,YS_119,YS_2}, quantum decision theory has been 
formulated for the case of a single decision maker taking a one-step decision. But in a 
society of decision makers, the agents exchange information between each other and 
can make multistep decisions, with their decisions varying with time due to the information 
exchange. A simplified situation, when all decision makers receive the same information
from outside, without mutual exchange, has been considered in \cite{YS_1,YS_29}. The 
necessity of developing, in the frame of QDT, a model describing the multistep 
decision-making procedure, taking into account mutual exchange of information between 
social agents, has been discussed in Ref. \cite{YS_2}.   

In the present paper, we suggest a principally important extension of QDT allowing 
for the description of dynamical collective effects. The main features, distinguishing
this paper from the previous publications, are as follows.

\begin{itemize}

\item
The considered system is a society of several decision makers, and not just a single
decision maker. 

\item
Each agent of the society is a decision maker transforming the information, received 
from other agents, according to quantum decision theory.

\item
At each decision step, every agent generates an outcome that is a probability 
distribution over a given set of prospects.

\item
Agent interactions are not parameterized by a fixed interaction matrix, but are 
characterized by an information functional over the probability distribution 
quantifying the amount of information gained from other agents.

\item
The system is not static, but dynamical, the information functional and generated
probability distributions are functions of time.

\item
To describe a realistic situation, the information exchange is not simultaneous, 
but delayed. This means that, if the agents at time $t$ generate probability 
distributions $\{p_j(t)\}$, then the probability distribution that follows, which 
is caused by the information exchange, is generated at a delayed time $t + \tau$.

\item
The usage of the theory is illustrated by a concrete example involving the dynamic 
disjunction effect. The predicted numerical results are in very good agreement with 
empirical data.

\end{itemize}

The proposed system is not a set of simple quantum devices representing players,
as those considered in quantum games 
\cite{Meyer_30,Eisert_31,Piotrowski_32,Landsburg_33,Guo_34}, but rather a collection 
of complex subjects exchanging information, each representing a quantum intelligence. 
Therefore the society of decision makers, acting according to the rules of QDT, 
is a kind of {\it collective quantum intelligence}, or superintelligence. 

Such a collective quantum intelligence can describe any ensemble of agents 
generating probability distributions according to the rules of QDT. This can be 
a human society making decisions on complex problems under uncertainty. This can 
also be a set of quantum computers, or a complex artificial intelligence composed 
of parts, each of which being itself an artificial quantum intelligence, that is, 
a kind of superbrain.  

The behavioral model we develop is principally new. The interaction structure 
of a multi-agent system, with agents interacting through the exchange of information 
occurring by quantum rules, has never been considered before to the best of our 
knowledge.

The organization of this article is as follows. Section 3 explains the basic ideas
characterizing the interaction structure of agents in a multi-agent society, where the 
interactions are due to the information exchange. In Sec. 4, we very briefly 
summarize the main ingredients determining the probability measure of a single 
given agent, which is based on quantum decision theory. Section 5 presents an 
extension of the theory of quantum decision making to the case of interacting 
agents forming a society. In these first sections, we introduce the notations
and definitions that are necessary for the following considerations. Section 6 treats 
the case of agents with long-term memory. Section 7 considers the case of agents 
with reconstructive memory, while Sec. 8 studies the situation of agents with 
short-term memory. In Section 9, we show that the society of decision makers, 
acting by the rules of QDT, can be interpreted as a novel type of networks -- 
a {\it quantum information network} or {\it quantum intelligence network}. To illustrate
the practical usage of the approach, in Sec. 10, we consider a concrete example of 
dynamic decision making. We analyze the dynamic disjunction effect, predicting the 
behavior of decision makers both, at the initial stage as well as in the long run, 
when there decisions converge to a common consensus. Our numerical predictions 
are in very good agreement with empirical data. Section 11 concludes.

\section{Interaction Structure for Agents Exchanging Information}

Here we explain the main ideas of how the interaction structure between agents 
exchanging information is organized.

Let us consider decision making in a society of $N$ agents choosing between $N_L$
prospects (lotteries) $\pi_n$, with $n = 1,2,\ldots,N_L$. At time $t$, a $j$-th agent 
generates a probability measure $\mathbb{P}_j(t) \equiv \{ p_j(\pi_n,t) \}$ giving the 
probabilities with which the prospects are to be chosen. As a result of the human 
decision making duality, the probability measure consists of two parts, a rational 
part quantifying the utility of the considered prospects and forming a classical 
probability measure $\mathbb{F}_j(t) \equiv \{ f_j(\pi_n,t) \}$ and a quantum part   
$\mathbb{Q}_j(t) \equiv \{ q_j(\pi_n,t): \; n = 1,2,\ldots,N_L \}$ characterizing 
irrational emotions and subconscious feelings, making a prospect subjectively 
attractive or not for an agent. 

Explicit definitions required for the practical usage will be given in the following 
sections. Meanwhile, it is sufficient to remember that the probability measure, 
$\mathbb{P}_j(t)$ generated by each agent, is the union of $\mathbb{F}_j(t)$
and $\mathbb{Q}_j(t)$.

The exchange of information between the agents implies that each agent receives
information on the probabilities generated by other agents. Thus the information 
exchange between two agents is represented by a directed graph in Fig. 1, showing 
that the first agent receives information from the second agent, and the latter, from 
the first one. For simplicity, this graph can be represented as in Fig. 2. For three agents,
their interactions through information exchange can be shown as in Fig. 3.  Similarly,
the interaction scheme can be extended to many agents.

\begin{figure}[!t]
\vspace{9pt}
\centering
\includegraphics[width=6cm]{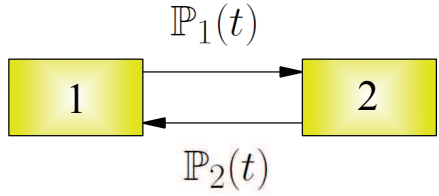} 
\caption{Information exchange between two agents.
}
\label{fig:Fig.1}
\end{figure}

\begin{figure}[!t]
\vspace{9pt}
\centering
\includegraphics[width=6cm]{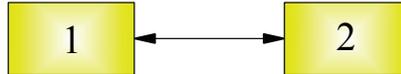} 
\caption{Simplified scheme of interaction through information 
exchange between two agents.
}
\label{fig:Fig.2}
\end{figure}

\begin{figure}[!t]
\vspace{9pt}
\centering
\includegraphics[width=6cm]{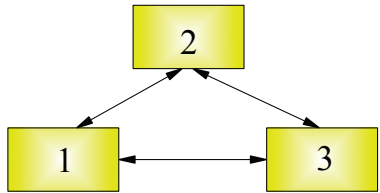} 
\caption{Scheme of interaction through information 
exchange between three agents.
}
\label{fig:Fig.3}
\end{figure}

After receiving information from other members of the society, each agent generates
the new probability measure that becomes available to other agents, and so on. 
Schematically, the dynamics of decision making with information exchange between 
two agents is shown in Fig. 4. The temporal variation of the probability measure   
$\mathbb{P}_j(t)$ includes the evolution of two parts, the classical probability measure 
 $\mathbb{F}_j(t)$, describing the utility or reward, whose dynamics is governed by the 
utility maximization, and the dynamics of the quantum part  $\mathbb{Q}_j(t)$ characterizing
the variation of emotions. Generally, the two types of dynamics do not need to be 
necessarily connected. Recall that a number of experimental studies  
\cite{Charness_66,Blinder_67,Cooper_68,Charness_69,Charness_70,Chen_71,
Charness_72,Kuhberger_73} demonstrated the evolution of behavioral effects,
caused by information exchange, under a fixed utility.

\begin{figure}[!t]
\vspace{9pt}
\centering
\includegraphics[width=8cm]{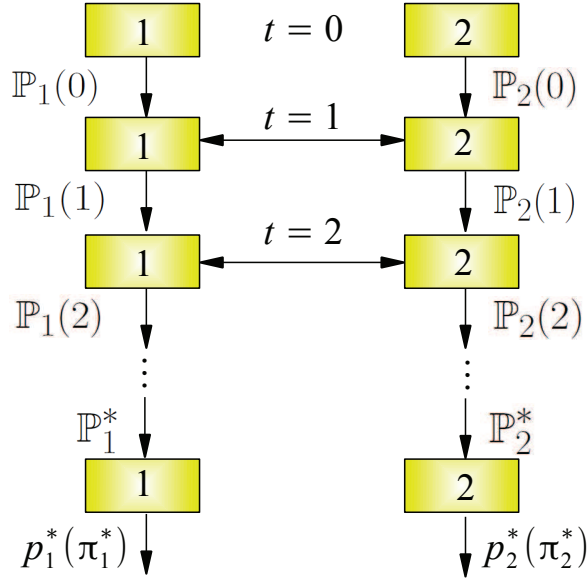} 
\caption{Dynamics of multistep decision making
governed by information exchange between two agents. The last step
represents the asymptotic convergence to fixed values of the probabilities that
contribute to the stationary values of the probabilities for a specific agent to choose a given prospect.
}
\label{fig:Fig.4}
\end{figure}

The convergence property of the decision making dynamics depends on the specification
of the agents' memory. When the process converges, then a $j$-th decision maker comes to
a stationary probability measure $\mathbb{P}_j^*$ composed of the set of probabilities
$p_j^*(\pi_n)$, each of which implies the probability of choosing a prospect $\pi_n$. 
Comparing these probabilities defines the optimal, for the $j$-th agent, prospect 
$\pi_j^*$ for which the probability is maximized: $p_j^*(\pi_j^*) = \max_n \{p_j^*(\pi_n)\}$.
As will be shown below, for short-term memory, the decision making process may become
not convergent.

\section{Probability Measure Generated by a Single Agent}

The theory of quantum probability distributions, generated by separate agents, 
has been developed in the previous papers 
\cite{YS_1,YS_2,YS_35,YS_36,YS_37,YS_38,YS_39}. The mathematical techniques 
of the theory are equally applicable for describing quantum measurements as well 
as quantum decision making \cite{YS_14,YS_15,YS_35}. As far as all mathematical 
details have been thoroughly exposed in our previous publications, here we only briefly 
recall the main points of the theory and introduce notations that will be necessary for 
the generalization in the next sections to the case of multi-agent systems.

The information processing by a single agent proceeds as follows. An agent 
receives information characterized by different types of events $A_n$ and $B_\alpha$, 
labelled by the indices $n$ and $\alpha$, and represented by quantum states,
\be
\label{1}
A_n \ra | n \rgl \; , \qquad B_\al \ra | \al \rgl
\ee
in the corresponding Hilbert spaces
\be
\label{2}
 \cH_A = {\rm span}\{ | n\rgl \} \; , \qquad  \cH_B = {\rm span}\{ | \al \rgl \} \; .
\ee
The difference between the events is in their degree of testability. The events 
$A_n$ are operationally testable, allowing for their unambiguous observation or 
measurement. In contrast, the events $B_\alpha$ are not operationally testable, 
being characterized by random amplitudes $b_\alpha$. The set of the random 
events $B = \{B_\alpha\}$ is an inconclusive event \cite{YS_13,YS_14,YS_15} 
represented by a state in the Hilbert space $\mathcal{H}_B$,
\be
\label{3}
 B = \{ B_\al \} \ra | B \rgl = \sum_\al b_\al | \al \rgl \;  .
\ee
Generally, information is provided through composite events
\be
\label{4}
\pi_n =  A_m \bigotimes B \ra |\pi_n \rgl = | nB \rgl
\ee
represented by states in the Hilbert space
\be
\label{5}
\cH = \cH_A \bigotimes \cH_B \;   .
\ee
These composite events are termed prospects. Note that the states 
$\vert \pi_n \rangle$ are not necessarily orthonormalized.

The prospects induce the prospect operators
\be
\label{6}
 \hat P(\pi_n) \equiv | \pi_n \rgl \lgl \pi_n | \;  ,
\ee
which, in general, are not necessarily projectors, because the related prospect 
states may be not orthonormalized. The set of prospect operators forms a positive 
operator-valued measure \cite{YS_40,YS_41}. The family of all prospect 
operators is equivalent to the algebra of local observables in quantum theory.

An agent is associated with a strategic state $\hat{\rho}$ that is non-negative 
and normalized, such that ${\rm Tr} \hat{\rho} = 1$, with the trace operation 
over space (\ref{5}). The agent generates the prospect probabilities
\be
\label{7}
  p(\pi_n) = {\rm Tr}\hat\rho \; \hat P(\pi_n)
\ee
forming a probability measure, so that
\be
\label{8}
 \sum_n p(\pi_n) =  1 \; , \qquad 0 \leq p(\pi_n) \leq 1 \;  .
\ee
A prospect probability consists of two parts,
\be
\label{9}
 p(\pi_n) = f(\pi_n) + q(\pi_n) \;  ,
\ee
where the first term
\be
\label{10}
 f(\pi_n) = \sum_\al |\; b_\al\;|^2 \lgl n\al | \; \hat\rho\; | n\al \rgl
\ee
is positive-definite, while the second term
\be
\label{11}
 q(\pi_n) = \sum_{\al\neq \bt}  b^*_\al b_\bt \lgl n\al | \; \hat\rho\; | n\bt \rgl
\ee
is not positive defined. The appearance in the probability of a not 
positive-defined term is due to the quantum definition of the probability and 
the interference of inconclusive events. Such a quantum term would be absent 
in classical probability.

Quantum information processing has to include, as a particular case, classical 
processing. Therefore the quantum-classical correspondence principle \cite{Bohr_42} 
has to be valid. In our case, this requires that the quantum probability should 
reduce to the classical probability when the quantum term tends to zero:
\be
\label{12}
 p(\pi_n) \ra f(\pi_n) \; , \qquad  q(\pi_n) \ra 0 \;  .
\ee
Thus, the positive-definite terms play the role of classical probabilities, with the 
standard properties
\be
\label{13}
\sum_n f(\pi_n) =  1 \; , \qquad 0 \leq f(\pi_n) \leq 1 \;    .
\ee
Then the quantum terms satisfy the conditions
\be
\label{14}
 \sum_n q(\pi_n) =  0 \; , \qquad -1 \leq q(\pi_n) \leq 1 \;   .
\ee
Keeping in mind conditions (\ref{8}) and (\ref{13}), we see that the quantum 
term can be either negative, such that 
$$
- f(\pi_n) \leq q(\pi_n) \leq 0 \;  ,
$$
or positive, when
$$
0 \leq q(\pi_n) \leq 1 - f(\pi_n) \; .
$$
In QDT, the term $f(\pi_n)$ describes the utility of the prospect $\pi_n$,
which justifies to call it the utility factor, while $q(\pi_n)$ characterizes
the attractiveness of the prospect and is called the attraction factor. Two terms,
appearing in decision making, reflect the duality of the latter, including the 
logical conscious evaluation of the prospect utility and subconscious intuitive
estimation of its attractiveness. The properties of these terms and  
their practical determination have been described in detail in the previous 
papers \cite{YS_1,YS_119,YS_2}.   
   
In this way, the information processing by a single agent consists in generating 
the probability measure over the given information characterized by the set of 
prospects. In the case of social networks, the classical probability $f(\pi_n)$ 
can be defined by using a kind of Luce choice axiom \cite{Luce_43,Blavatskyy_44}. 
More generally, it can be defined as a minimizer of an information functional or 
by conditional entropy maximization \cite{YS_1,YS_2}. Note that unconditional entropy 
maximization may occur to be not sufficient for correctly defining probabilities,
in which case one has to respect additional constraints making the ensemble 
representative \cite{YS_1,YS_2,Yukalov_90}. 
 
The quantum term $q(\pi_n)$ depends on the amount of information $M$ received 
by the agent. In the absence of information, the initial value $q_0(\pi_n)$ can be 
random, although satisfying conditions (\ref{14}). After getting the amount of 
information $M$, the quantum term can be written \cite{YS_1,YS_29} as
\be
\label{15}
 q(\pi_n) = q_0(\pi_n) \exp(-M) \; .
\ee
In quantum decision theory, the quantum term characterizes the attractiveness of the 
considered prospects, and it is thus called the attraction factor. 
 
Upon the receipt of additional  information, the non-utility part of the probability measure 
generated by the  agent is reduced. In the previous papers \cite{YS_1,YS_29}, we have
considered the simple case where the amount of information $M$ is the same for 
all agents, being given by an external source or by a control algorithm. 
Now we shall study a more realistic situation, when the agents receive information by 
exchanging it with other members of the society.

\section{Generalization to Multiple Agents Exchanging Information}

\subsection{Arbitrary number of prospects}

Let us now generalize the procedure of information processing by a single agent 
to the case of many agents forming a society. Let the agents be enumerated by 
$j = 1,2,\ldots,N$. At the initial time, the information is presented through a set of 
prospects $\pi_n$ enumerated by $n = 1,2,\ldots,N_L$. The process of getting 
additional information requires time, so that the probability measure generated 
by the $j$-th agent is a function of time:
\be
\label{16}
 p_j(\pi_n,t) \equiv p_{jn}(t) \; .
\ee
Respectively, the utility and attraction factors are also functions of time:
\be
\label{17}
  f_j(\pi_n,t) \equiv f_{jn}(t) \; , \qquad  q_j(\pi_n,t) \equiv q_{jn}(t) \;  .
\ee
There always exist the normalization conditions for the probabilities,
\be
\label{18}
 \sum_{n=1}^{N_L} p_{jn}(t) =  1 \; , \qquad 0 \leq p_{jn}(t) \leq 1 \;  ,
\ee
for the utility factors,
\be
\label{19}
  \sum_{n=1}^{N_L} f_{jn}(t) =  1 \; , \qquad 0 \leq f_{jn}(t) \leq 1 \;  ,
\ee
and for the attraction factors,
\be
\label{20}
  \sum_{n=1}^{N_L} q_{jn}(t) =  0 \; , \qquad -1 \leq q_{jn}(t) \leq 1 \;  .
\ee

In realistic situations, the probability measure is not immediately generated by 
the $j$-th agent but after a delay time $\tau$. Taking this into account and 
measuring time in units of $\tau$, we can write
\be
\label{21}
 p_{jn}(t+1) =  f_{jn}(t) +  q_{jn}(t)  \;    .
\ee
The first term is a utility factor, whose value is prescribed by the objective 
utility of the problem, hence assumed to be defined by prescribed rules 
\cite{YS_1,YS_119,YS_2}. The second term has been shown \cite{YS_1,YS_29} 
to be a function of the information measure $M_j(t)$ quantifying the amount 
of information received by the $j$-th agent until time $t$,
\be
\label{22}
 q_{jn}(t) = q_{jn}(0)\exp\{ -M_j(t) \} \;   .
\ee
The initial value $q_{jn}(0)$ is a random quantity satisfying the above 
normalization conditions.

It is worth stressing the importance of taking into account the delay in receiving the 
information, which makes the consideration realistic, since in real life acquiring 
information always requires finite time. On the other side, the occurrence of time 
delay can essentially change the dynamics of multi-agent systems.   

The total information received by the $j$-th agent until time $t$ can be represented 
as a sum
\be
\label{23}
 M_j(t) = \sum_{k=1}^t \hat{\vp}_j(t,k) \mu_j(k) \;  ,
\ee
where $\mu_j(k)$ is the information gained by the $j$-th agent at the $k$-th time step 
and $\hat{\vp}_j(t,k)$ is a memory operator to be specified below, which defines 
how much of the information at time $k$ is retained at the later time $t > k$.

Information measures can be chosen in different ways, for instance as transfer 
entropy \cite{Schreiber_47,Lizier_48,Smirnov_49,Sun_50} or in the standard form 
of the Kullback-Leibler relative information \cite{Kullback_51,Kullback_52,Vedral_53}. 
We prefer the latter approach. Then, the information gain for the $j$-th agent concerning 
the $n$-th prospect can be written as
\be
\label{24}
 \mu_j(k) = \sum_{n=1}^{N_L} \; p_{jn}(k) \ln\frac{p_{jn}(k)}{h_{jn}(k)} \; ,
\ee
where 
\be
\label{25}
 h_{jn}(k) = \frac{1}{N-1} \sum_{i(\neq j)} \; p_{in}(k)
\ee
is the average probability for the $n$-th prospect over all agents of the society, 
except the $j$-th agent.

\subsection{Case of two prospects}

The problem simplifies when there are only two prospects ($N_L = 2$). This is 
actually a situation that is very often met, corresponding to choosing between 
just two alternatives, when deciding on ``yes" or ``no". Then, keeping in mind the 
normalization conditions (\ref{18}) to (\ref{20}), it is sufficient to consider only 
one of the prospects, say $\pi_1$, thus simplifying the notation for the 
first-prospect probability,
\be
\label{26}
  p_j(t) \equiv p_{j1}(t) = p_j(\pi_1,t) \; ,
\ee
since the second-prospect probability is
\be
\label{27}
  p_j(\pi_2,t) \equiv p_{j2}(t) = 1-  p_j(t) \;  .
\ee
Similar simplified notations can be used for the utility factors,
$$
 f_j(t) \equiv f_{j1}(t) = f_j(\pi_1,t) \; , 
$$
\be
\label{28}
f_j(\pi_2,t) \equiv f_{j2}(t) = 1-  f_j(t) \; ,
\ee
and the attraction factors,
$$
 q_j(t) \equiv q_{j1}(t) = q_j(\pi_1,t) \; , 
$$
\be
\label{29}
 q_j(\pi_2,t) \equiv  q_{j2}(t)  = - q_j(t) \; ,
\ee
each associated with the $j$-th agent. 

The utility factors, reflecting the basic utility of prospects, can be treated 
as time-independent, which we shall use in what follows:
\be
\label{30}
 f_j(t) = f_j(0) = f_j \;  .
\ee

To proceed further, we need to specify the memory operator characterizing the
type of memory typical of the considered agents.

\subsection{Types of memory}

Different types of memory have been  classified in psychology and neurobiology
\cite{Baddeley_75,Cowan_76,Schacter_77}.  For the purpose of the present 
consideration, we need to distinguish the types of memory defining in different 
ways the temporal behavior of the information measure $M_j(t)$. It is possible 
to distinguish three qualitatively different kinds of temporal memory:

(i) {\it Long-term memory}, when there is no memory attenuation and all information, 
gained in the past, is perfectly retained. This implies that the memory operator
is identical to unity, $\hat{\varphi}_j(t,k) = 1$. 

(ii)  {\it Reconstructive memory}, when only the closest events are kept in mind,
while all previous temporal blanks are filled in by reconstructing the past by 
analogy with the present time \cite{Bartlett_78,Burgess_79,Schacter_80}.
This kind of memory can be represented by the action of the memory operator
$\hat{\varphi}_j(t,k) \mu_j(k) = \mu_j(t)$.

(iii) {\it Short-term memory}, when the memory of the past quickly attenuates.
In the very short-term variant, only the memory from the last step is retained,
while nothing is remembered from the previous temporal steps. This assumes
the local memory operator $\hat{\varphi}_j(t,k) = \delta_{kt}$, which defines the
Markov-type memory in decision process.
       
Below, we analyze in turn the behavior of the agents possessing these three 
principally different types of memory.

\section{Agents with Long-Term Memory}

Let us consider the limiting case of long-term memory corresponding to a 
non-decaying memory characterized by the memory operator 
$\hat{\varphi}_j(t,k) = 1$. As a consequence, the total accumulated information 
(\ref{23}) is the sum of the information gains at each temporal step,
\be
\label{31}
M_j(t) = \sum_{k=1}^t \mu_j(k) \;  .
\ee

For simplicity, we again consider the case of two prospects ($N_L = 2$). Moreover, 
we assume that agents can be divided into two groups, so that all agents 
within each group have the same initiation conditions. This amounts to consider 
a situation with two ``group-agents", or ``superagents" representing two agent groups, 
exchanging information with each other. We thus have the equations for the probabilities:
\be
\label{32}
 p_j(t+1) = f_j + q_j(t) \;  ,
\ee
with $j = 1, 2$. The utility factors $f_j$ are assumed to be constants characterizing 
the intrinsic utility of the prospects for the agents. Generally, the agents of a society are
heterogeneous, and thus the values of $f_j$ are different for different agents.  
The attraction factors vary with time as
\be
\label{33}
 q_j(t) = q_j(0) \exp\{ - M_j(t) \} \;  ,
\ee
where $q_j(0)$ are the chosen initial conditions. The temporal variation 
is influenced by the accumulated information defined in Eq. (\ref{31}). 
The information gain (\ref{24}) of the $j$-th agent, at a $k$-th step, reads as
\be
\label{34}
\mu_j(k) = p_j(k) \ln \frac{p_j(k)}{p_i(k)} +
[ 1 - p_j(k)] \ln \frac{1-p_j(k)}{1-p_i(k)}  \;  ,
\ee
with $j \neq i$. The initial conditions for the probabilities are
\be
\label{35}
 p_j(0) = f_j + q_j(0) \;  .
\ee
We also assume that there is no additional information at the beginning, when 
$t = 0$, so that $M_j(0) = 0$. Time varies in discrete steps as $t = 0, 1, 2, \ldots$. 
Note that, for any $k \geq 0$, the information gains $\mu_j(k)$ defined in (\ref{34}) are 
always positive. It follows that the total information gain $M_j(t) > 0$ is 
positive for $j = 1,2$ and $t \geq 0$.

Analyzing the behavior of the probabilities for varying initial conditions, 
we find that there are two qualitatively different types of solutions, depending 
on whether there exists an initial conflict between the utility factors and 
probabilities, or not. The existence or absence of an initial conflict is 
understood in the following sense. 

(i) There is no initial conflict when either
\be
\label{36}
f_1 > f_2 \; , \qquad p_1(0) > p_2(0) \qquad ({\rm no \; conflict}) \; ,
\ee
or
\be
\label{37}
 f_1 < f_2 \; , \qquad p_1(0) < p_2(0) \qquad ({\rm no \; conflict}) \; .
\ee
This implies that, at the initial time, the agents already prefer the prospect 
with a larger utility. 

(ii) There exists an initial conflict when either
\be
\label{38}
  f_1 < f_2 \; , \qquad p_1(0) > p_2(0) \qquad ({\rm conflict}) \; ,
\ee
or 
\be
\label{39}
 f_1 > f_2 \; , \qquad p_1(0) < p_2(0) \qquad ({\rm conflict}) \;  .
\ee
This occurs when, at the initial time, the agents prefer the less useful prospects.
Let us recall that such a conflicting choice very often occurs in decision 
making under uncertainty \cite{YS_1,YS_119,YS_2}. 

If there is no initial conflict, the probabilities $p_j(t)$ tend to the corresponding 
$f_j$, when $t \ra \infty$, as is shown in Fig. 5. But in the presence of an initial 
conflict, the numerical analysis of equations (\ref{32}) to (\ref{35}) shows that 
both probabilities tend to the common limit $p^*$, defined as
\be
\label{40}
p^* = \frac{f_1 q_2(0) - f_2 q_1(0)}{q_2(0) - q_1(0)} \; ,
\ee
within an accuracy of $10^{-3}$. This is illustrated in Fig. 6. 

It is worth stressing that the consensual limit (\ref{40}), under conflicting 
initial conditions (\ref{38}) or (\ref{39}), satisfies the inequalities 
$$
0 < p^* < 1 \; ,
$$
which is proved as follows.

Suppose that conditions (\ref{38}) are valid. Then, using (\ref{35}), we get
$$
q_2(0) - q_1(0) < f_1 - f_2 < 0 \; .
$$
In the case of conditions (\ref{39}), we find
$$
0 < f_1 - f_2 < q_2(0) - q_1(0) \; .
$$
In both these cases,
$$
0 < \frac{f_1-f_2}{q_2(0) - q_1(0)} < 1 \;  .
$$
Since $0 \leq p_j(0) \leq 1$ and $0 \leq f_j \leq 1$, with $-1 \leq q_j(0) \leq 1$, 
for any $C \in (0,1)$, we have
$$
0 < f_j + C q_j(0) < 1 \qquad ( 0 <  C < 1 ) \;   .
$$
Substituting here
$$
C = \frac{f_1-f_2}{q_2(0) - q_1(0)} \; ,
$$
we come to the result $0 < p^* < 1$.

\begin{figure*}[!t]
\vspace{9pt}
\centerline{
\hbox{ \includegraphics[width=7.5cm]{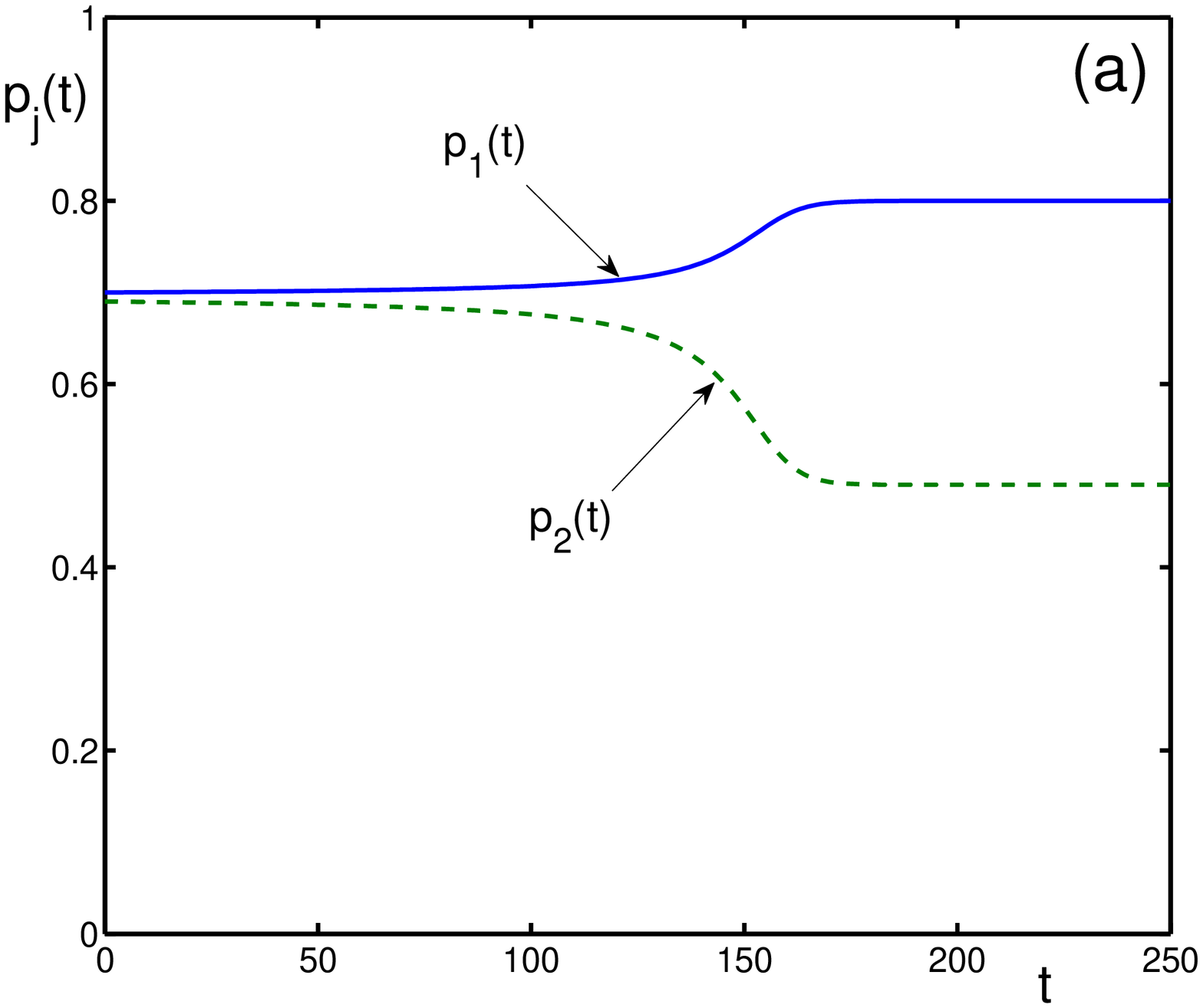}
\hspace{1.5cm}
\includegraphics[width=7.5cm]{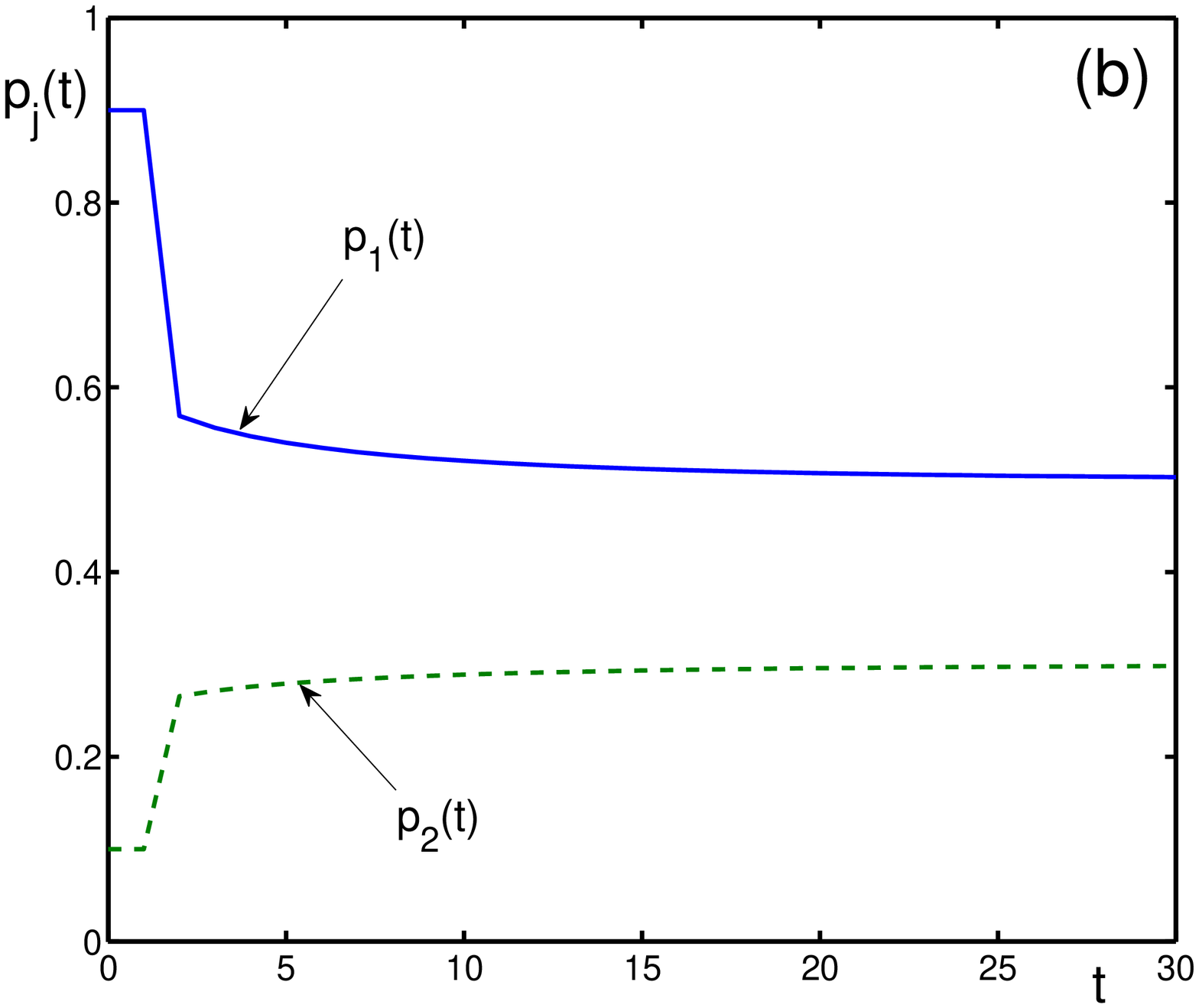} } }
\caption{Long-term memory under no conflict. Dynamics of the probabilities
$p_1(t)$ for superagent 1 (solid line) and $p_2(t)$ for superagent 2 (dashed line), when there
is no initial conflict, for two cases of close and rather different initial probabilities:
(a) initial conditions are $f_1 = 0.8 > f_2 = 0.49$, with $q_1(0) = -0.1$ and $q_2(0) = 0.2$,
so that the initial probabilities $p_1(0) = 0.7 > p_2(0) = 0.69$ are close to each other;
(b) initial conditions are $f_1 = 0.5 > f_2 = 0.3$, with $q_1(0) = 0.4$ and $q_2(0) = -0.2$,
so that the initial probabilities $p_1(0) = 0.9 > p_2(0) = 0.1$ are far from each other.
In both cases, there is no conflict and the probabilities $p_j(t)$ tend to their respective
limits $f_j$.
}
\label{fig:Fig.5}
\end{figure*}

\begin{figure*}[!t]
\vspace{9pt}
\centerline{
\hbox{ \includegraphics[width=7.5cm]{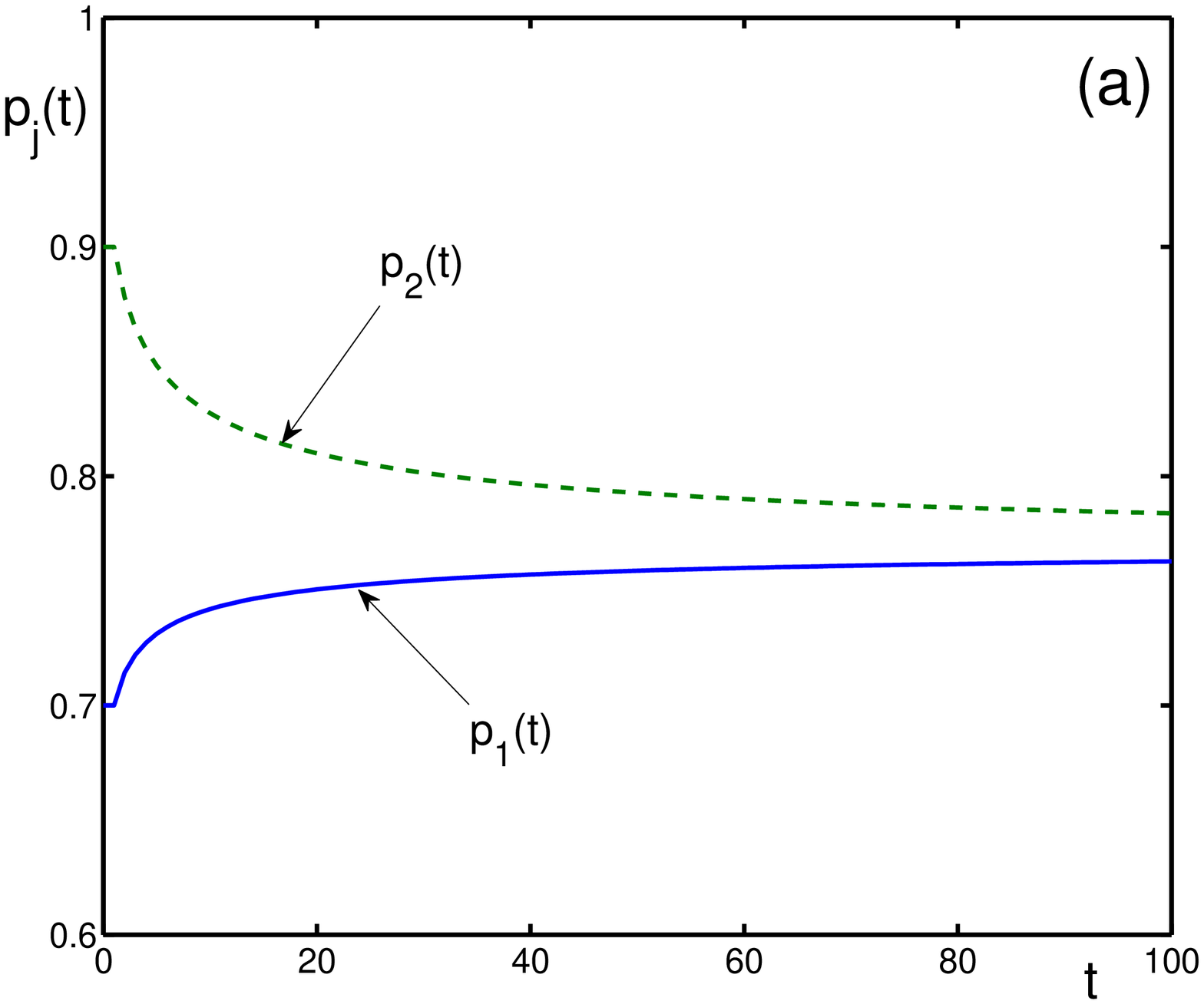}
\hspace{1.5cm}
\includegraphics[width=7.5cm]{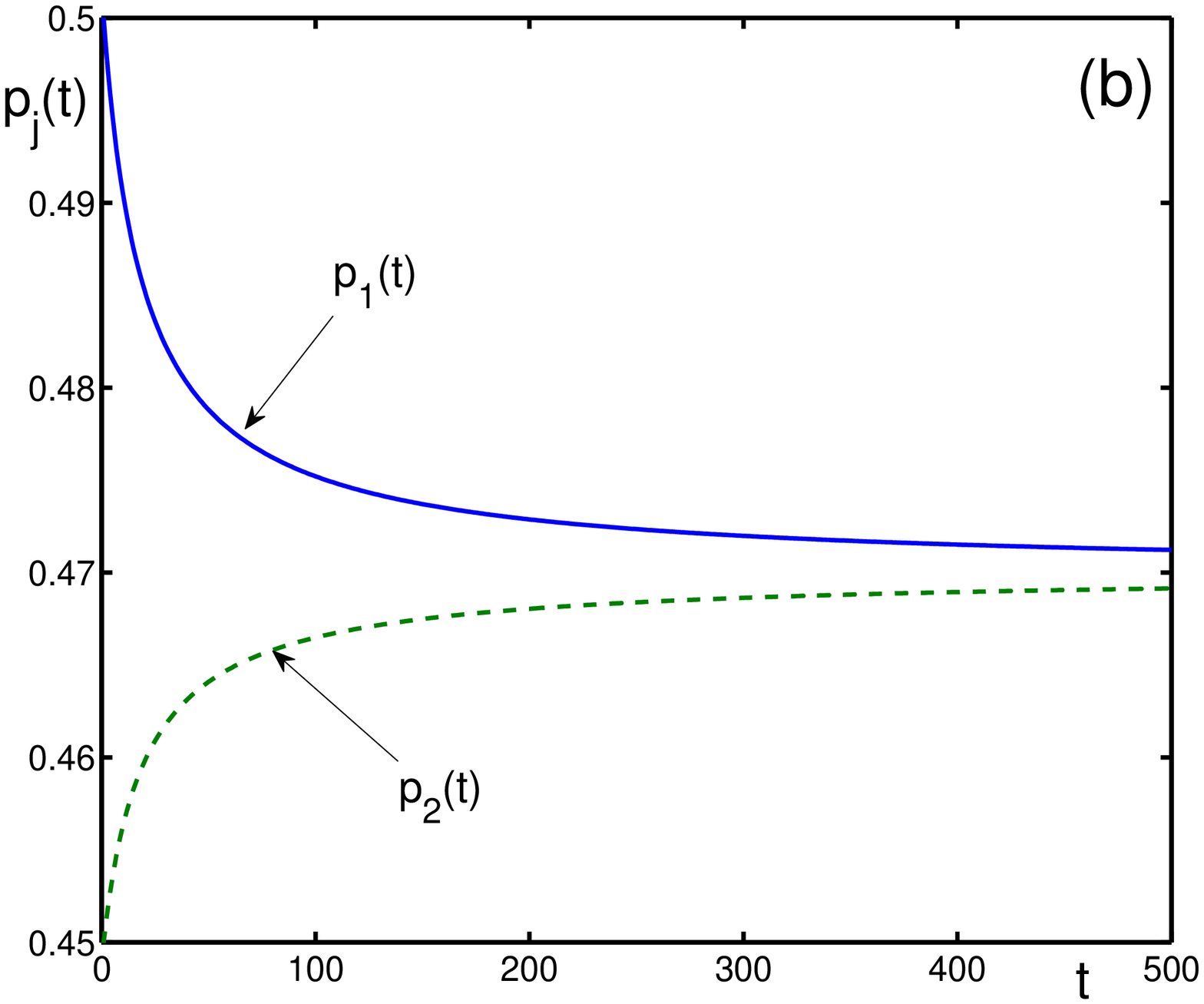} } }
\caption{Long-term memory under conflict. Dynamics of the probabilities $p_1(t)$
(solid line) and $p_2(t)$ (dashed line) in the presence of an initial conflict:
(a) initial conditions are $f_1=0.8>f_2=0.7$, with $q_1(0)=-0.1$ and $q_2(0)=0.2$, so that
$p_1(0)=0.7<p_2(0) =0.9$, the consensual limit is $p^* = 0.77$; (b) initial conditions are
$f_1=0.2<f_2=0.65$, with $q_1(0)=0.3$ and $q_2(0)=-0.2$, so that $p_1(0)=0.5>p_2(0)=0.45$,
the consensual limit is $p^* = 0.47$.
}
\label{fig:Fig.6}
\end{figure*}

\section{Agents with Reconstructive Memory}

Let us consider the situation corresponding to agents with reconstructive
memory, when they put large weight to the last information gain at time $t$, 
and use it to fill up  the blanks over all previous times. This implies the action 
of the memory operator $\hat{\varphi}_j(t,k) \mu_j(k) = \mu_j(t)$. Then the 
total information received by the $j$-th agent at time $t$ is
\be
\label{41}
M_j(t) =  \sum_{k=1}^t  \mu_j(t) = \mu_j(t)~ t  \;  .
\ee
Except for the total information (\ref{41}), all other formulas are the same 
as in the previous section.

The analysis shows that, for all initial conditions with different $f_j$ and any $q_j$, 
the probabilities $p_j(t)$ tend to their utility factors $f_j$,
\be
\label{42}
 \lim_{t\ra\infty} p_j(t) = f_j \;  .
\ee
However, this tendency can be of three types. If the initial conditions are 
not conflicting, in the sense of inequalities (\ref{36}) or (\ref{37}), the tendency 
can be either monotonic, as in Fig. 7 or with oscillations, as in Figs 8 and 9. 
But when the initial conditions are conflicting, in the sense of Eqs. (\ref{38}) 
or (\ref{39}), then there always appear oscillations at intermediate stages, 
as is shown in Figs. 10 and 11. In the marginal case, when $f_1$ and $f_2$ 
coincide, the oscillations last forever, as illustrated in Fig. 9. However, this regime is 
not stable and disappears under any small difference between the $f_j$'s, leading to
damped oscillations.
 
In the case where there is no conflict, the oscillations appear more regular while, 
in the case with an initial conflict, they look more chaotic. In order to understand better 
the oscillation characteristics, we calculate the local Lyapunov exponents, as is 
explained in Ref. \cite{Yukalov_54}. For this purpose, we define the multiplier 
matrix $\hat m(t) \equiv [ m_{ij}(t) ]$ with the elements
\be
\label{43}
m_{ij}(t) =  \frac{\dlt p_i(t+1)}{\dlt p_j(t)} =  \frac{\dlt q_i(t)}{\dlt p_j(t)} \; ,
\ee
whose eigenvalues are
\be
\label{44}
e_{1,2} = \frac{1}{2} \left [ {\rm Tr}\;\hat m \pm
\sqrt{({\rm Tr}\; \hat m)^2 - 4 {\rm det}\;\hat m } \right ] \;  ,
\ee
where
$$
{\rm Tr}\; \hat m \equiv m_{11} + m_{22} \; , \qquad
{\rm det}\; \hat m \equiv m_{11}m_{22} - m_{12}m_{21} \;   .
$$
Then for the local Lyapunov exponents we have
\be
\label{45}
  \lbd_{1,2}(t) \equiv \frac{1}{t} \; \ln | e_{1,2} | \; .
\ee
At the intermediate stage characterized by strong oscillations of the 
probabilities, at least one of the local Lyapunov exponents becomes 
transiently positive, thus, demonstrating local instability. But in the 
long run,  the dynamics is always stable since, at large $t$, the Lyapunov 
exponents are negative, as is seen from Figs. 7 to 11.

\begin{figure*}[!t]
\vspace{9pt}
\centerline{
\hbox{ \includegraphics[width=7.5cm]{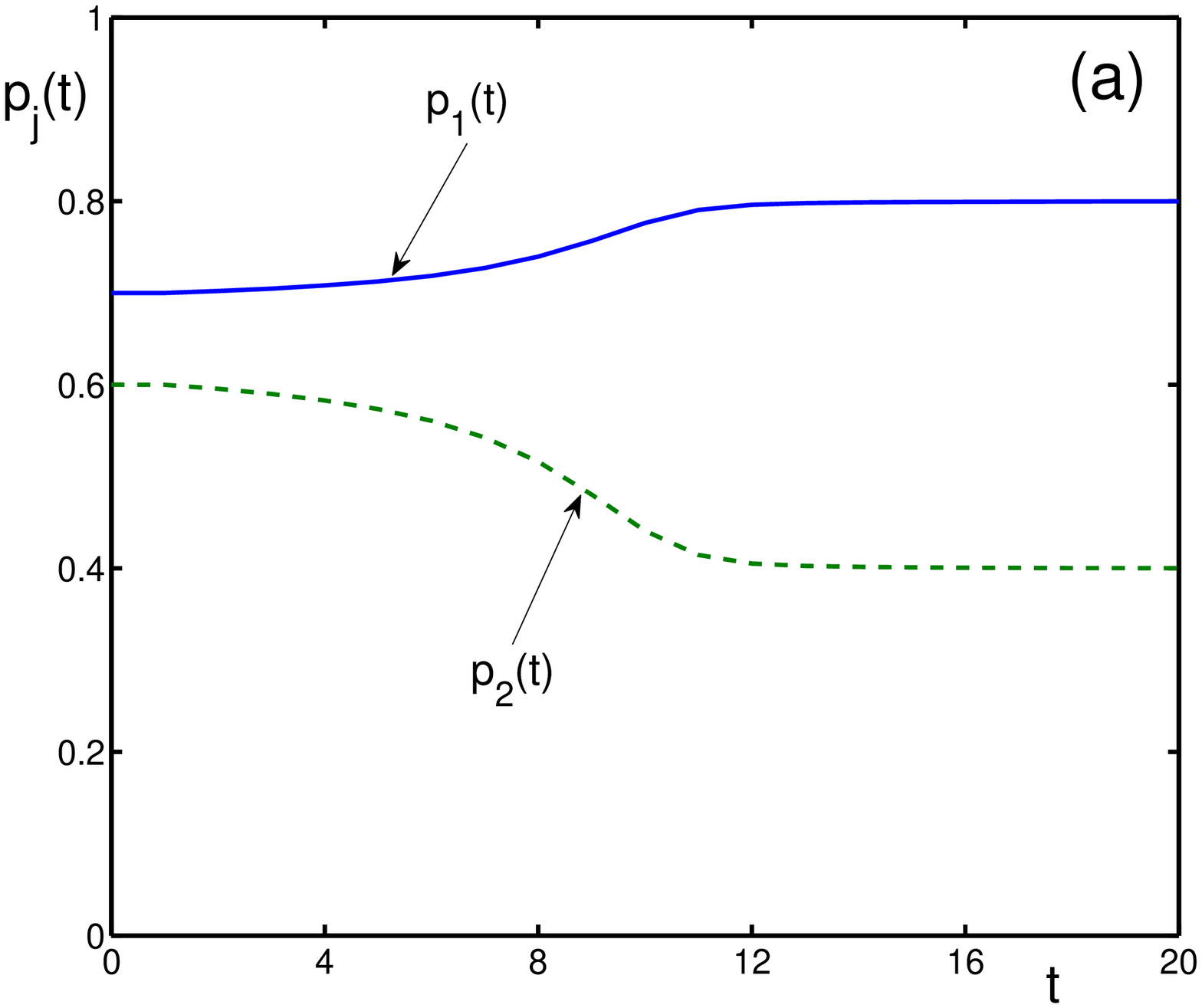}
\hspace{1.5cm}
\includegraphics[width=7.5cm]{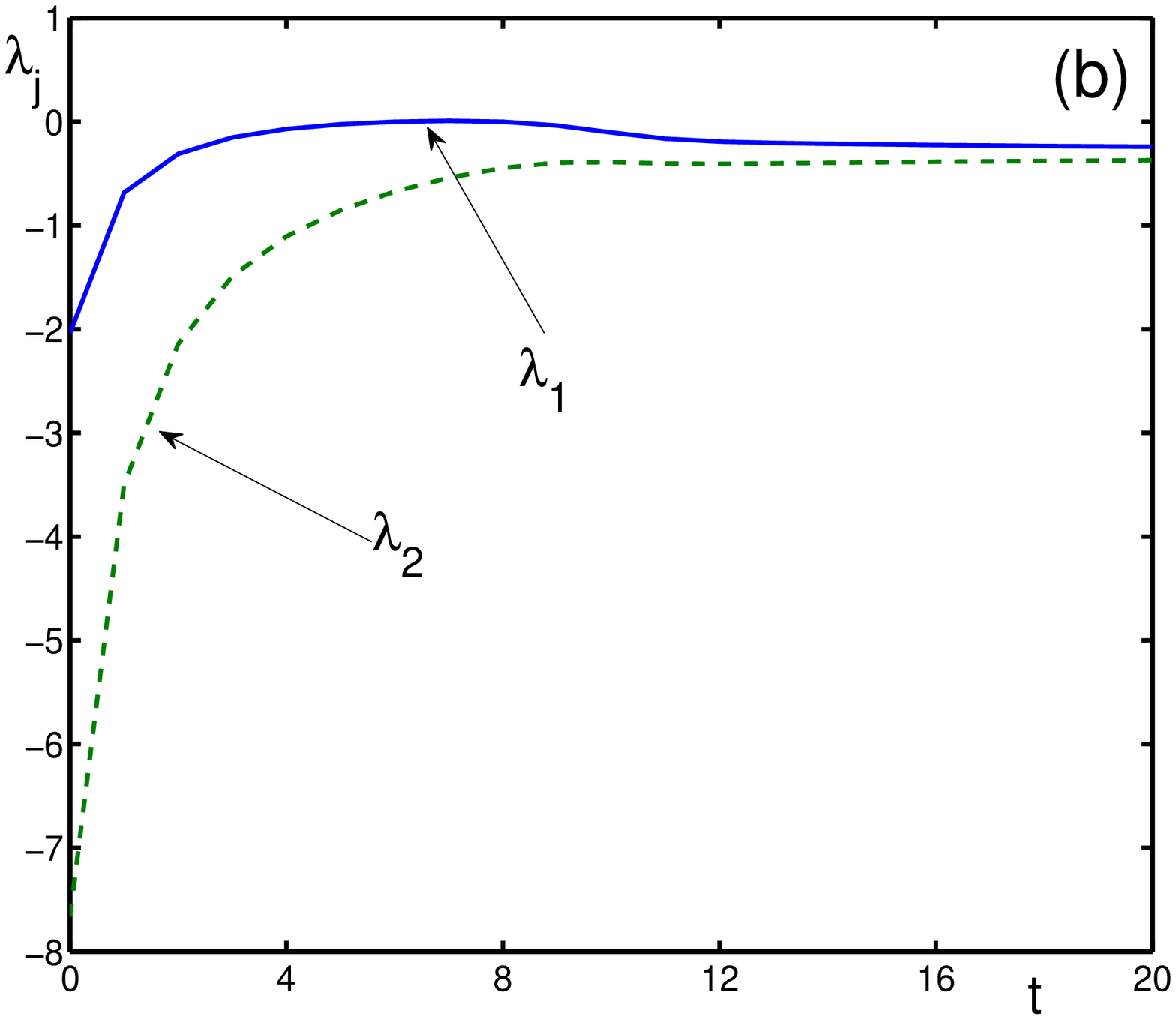} } }
\caption{Reconstructive memory under no conflict. Dynamics of the probabilities $p_j(t)$
and local Lyapunov exponents for the initial conditions $f_1=0.8>f_2=0.4$, with
$q_1(0)=-0.1$ and $q_2(0) = 0.2$, so that $p_1(0) = 0.7 > p_2(0) = 0.6$: (a) probabilities;
(b) local Lyapunov exponents.
}
\label{fig:Fig.7}
\end{figure*}

\begin{figure*}[!t]
\vspace{9pt}
\centerline{
\hbox{ \includegraphics[width=7.5cm]{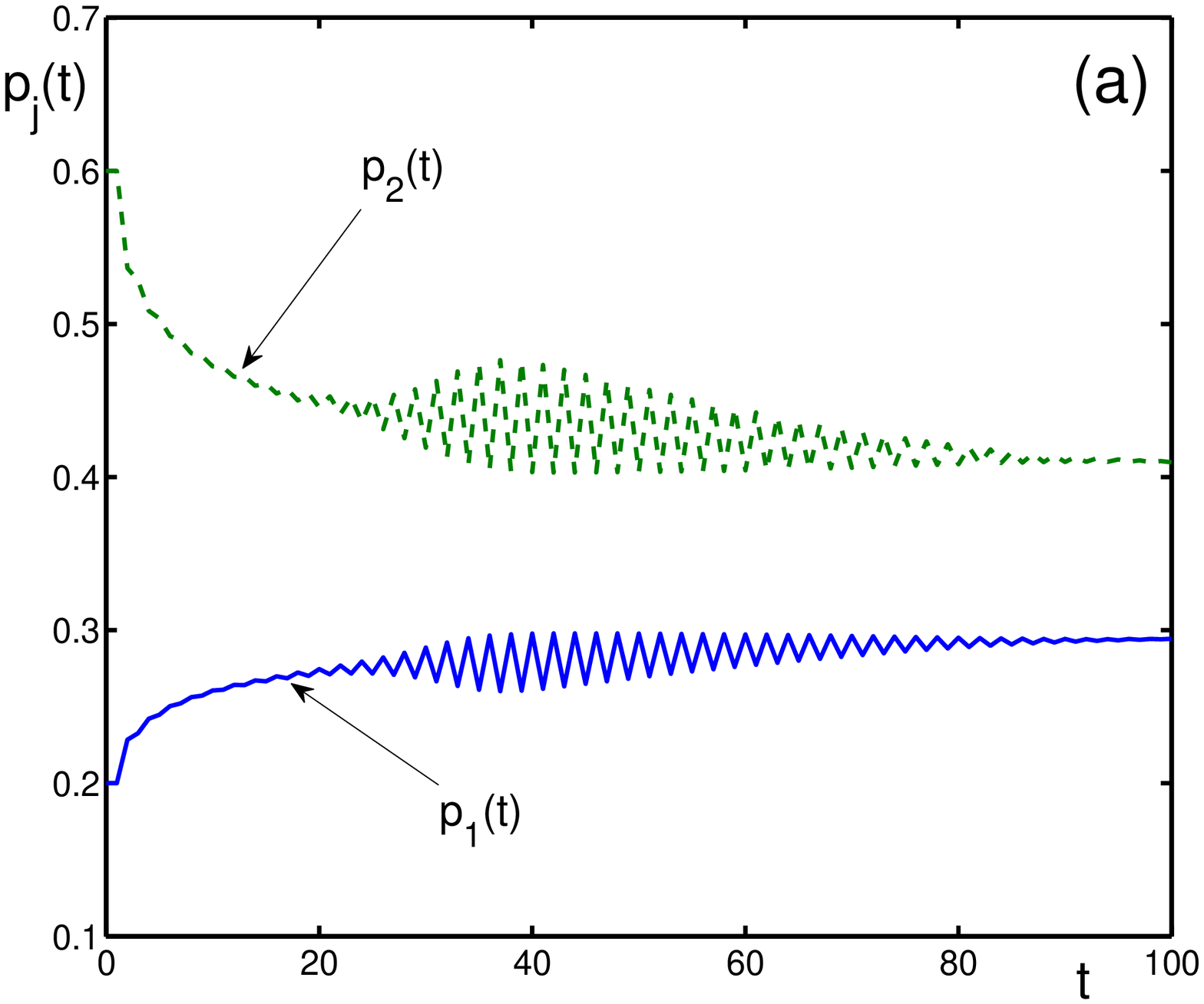}
\hspace{1.5cm}
\includegraphics[width=7.5cm]{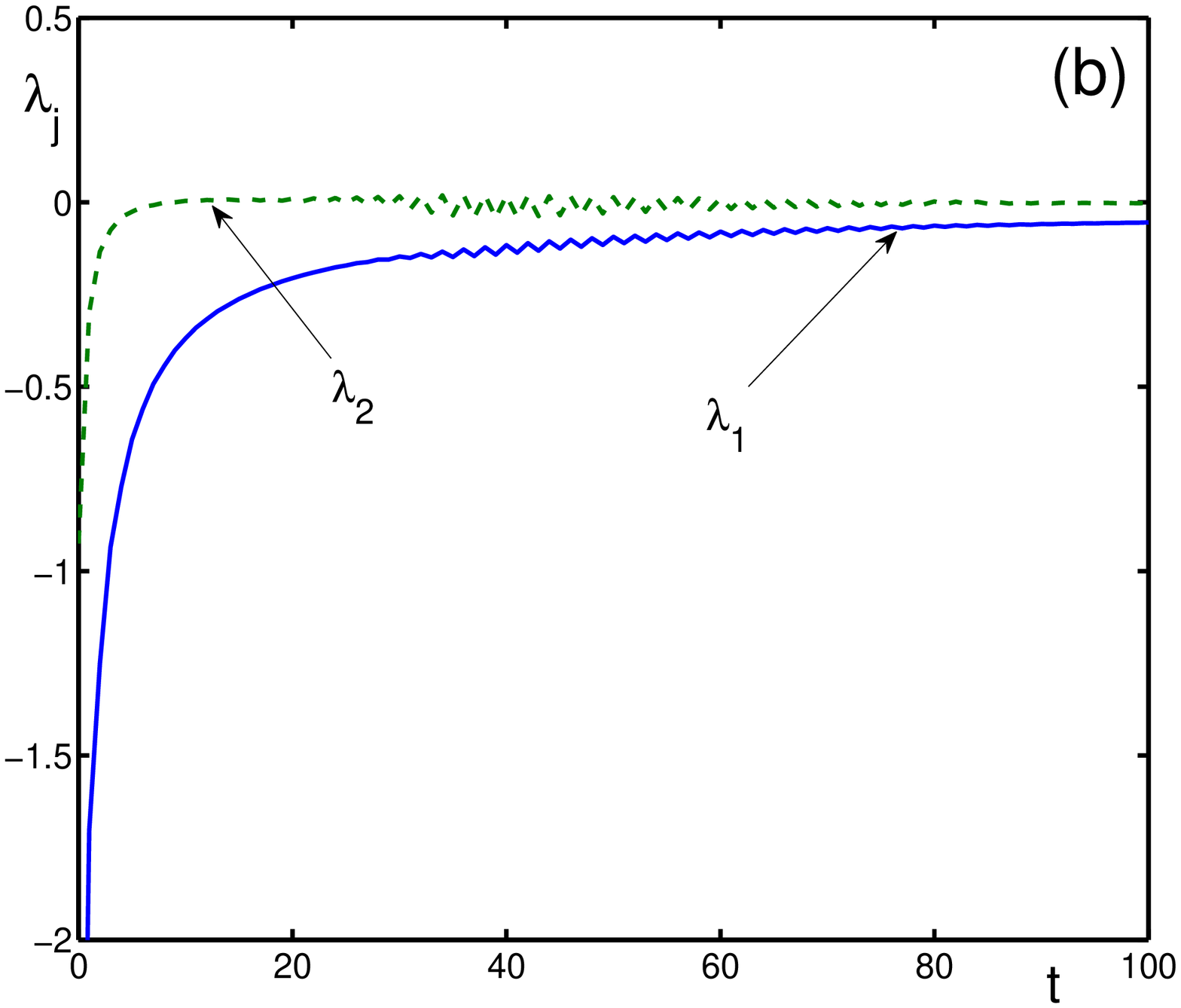} } }
\caption{Reconstructive memory under no conflict. Weak oscillations of the probabilities
(a) and smooth local Lyapunov exponents (b) for the initial conditions $f_1=0.3<f_2=0.4$,
with $q_1(0) = -0.1$ and $q_2(0) = 0.2$, so that $p_1(0) = 0.2 < p_2(0) = 0.6$.
}
\label{fig:Fig.8}
\end{figure*}

\begin{figure*}[!t]
\vspace{9pt}
\centerline{
\hbox{ \includegraphics[width=7.5cm]{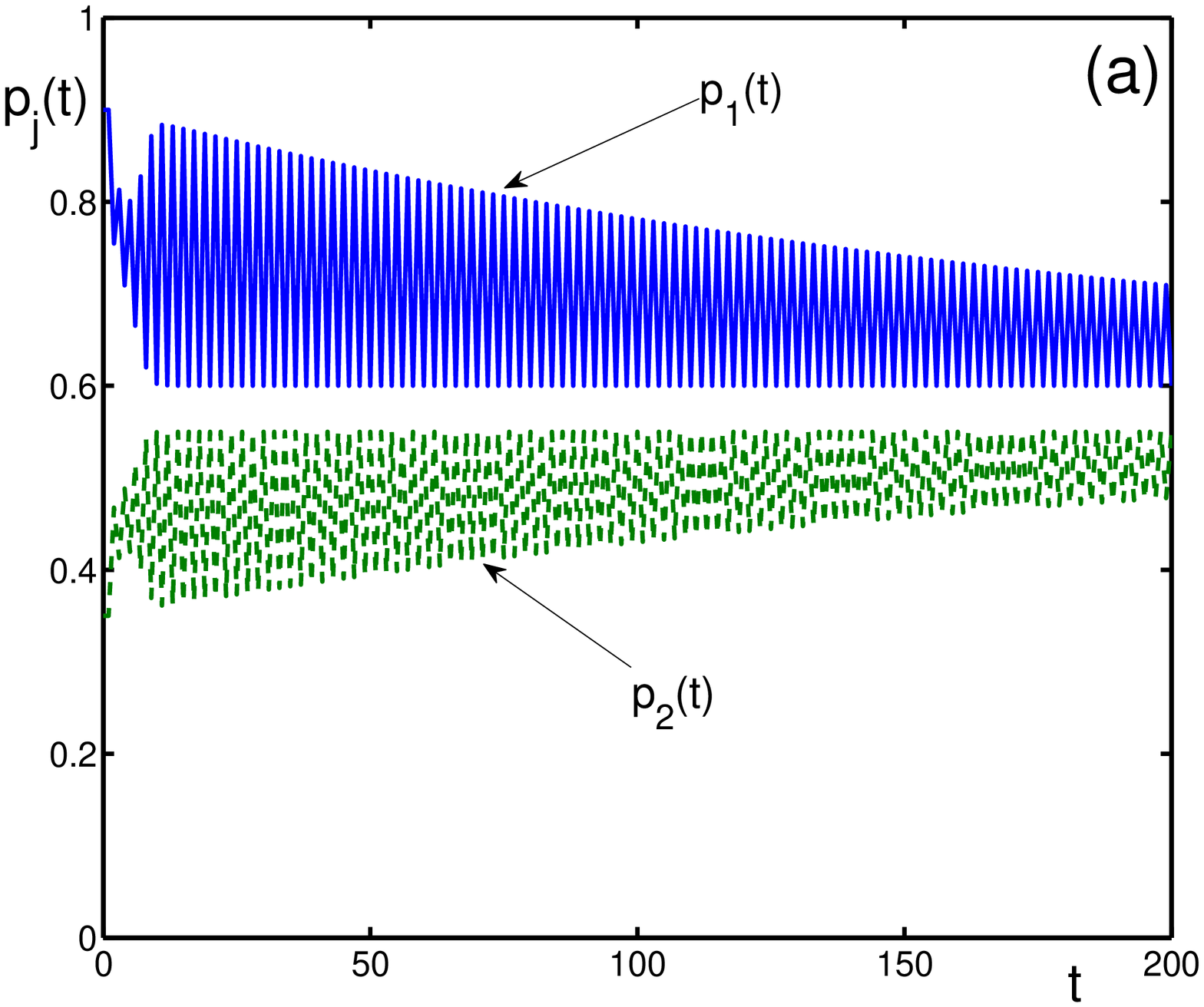}
\hspace{1.5cm}
\includegraphics[width=7.5cm]{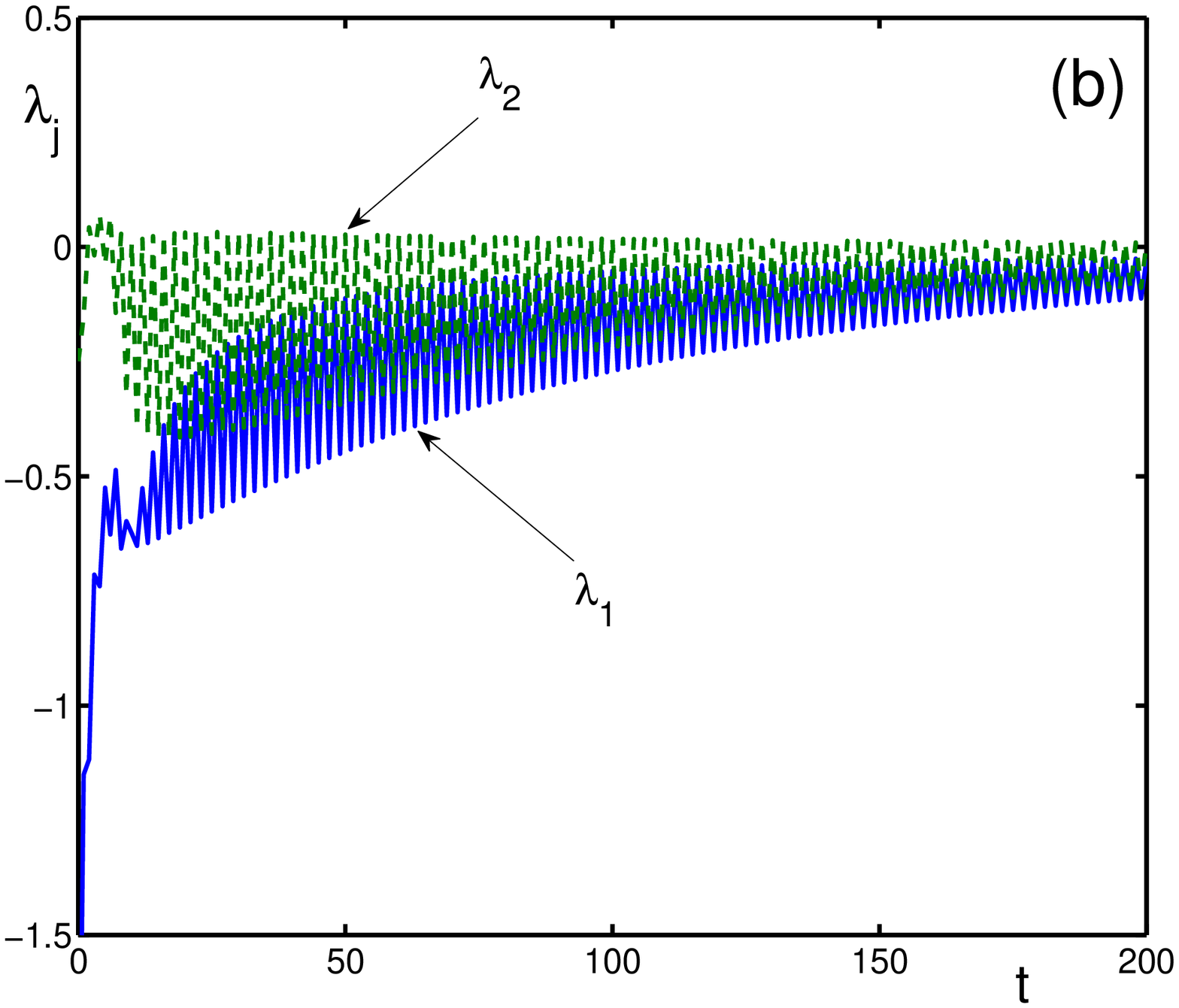} } }
\caption{Reconstructive memory under no conflict. Strong oscillations of the
probabilities (a) and oscillating local Lyapunov exponents (b) for the initial conditions
$f_1=0.6>f_2=0.55$, with $q_1(0)=0.3$ and $q_2(0)=-0.2$, so that $p_1(0)=0.9>p_2(0)=0.35$.
}
\label{fig:Fig.9}
\end{figure*}

\begin{figure*}[!t]
\vspace{9pt}
\centerline{
\hbox{ \includegraphics[width=7.5cm]{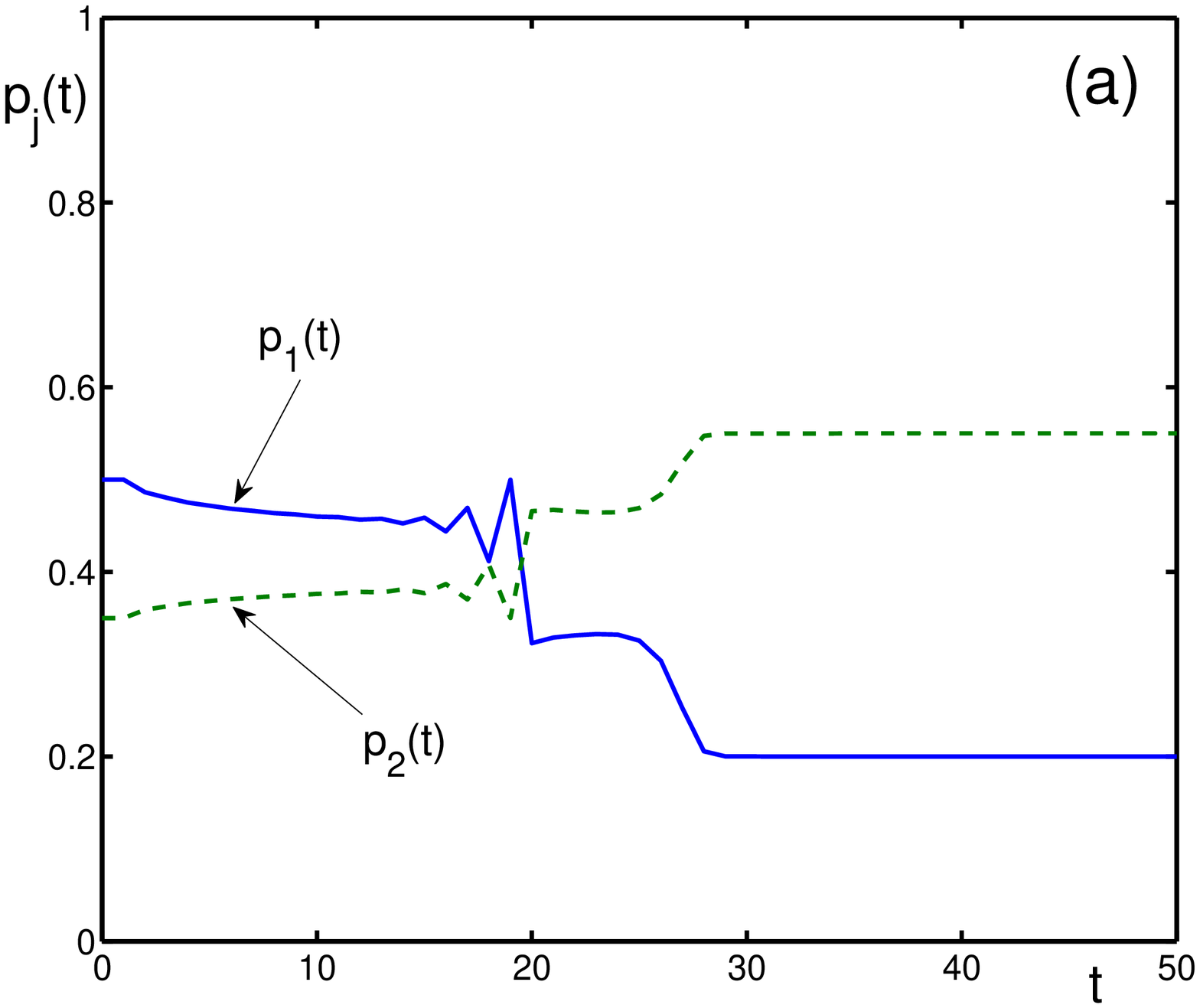}
\hspace{1.5cm}
\includegraphics[width=7.5cm]{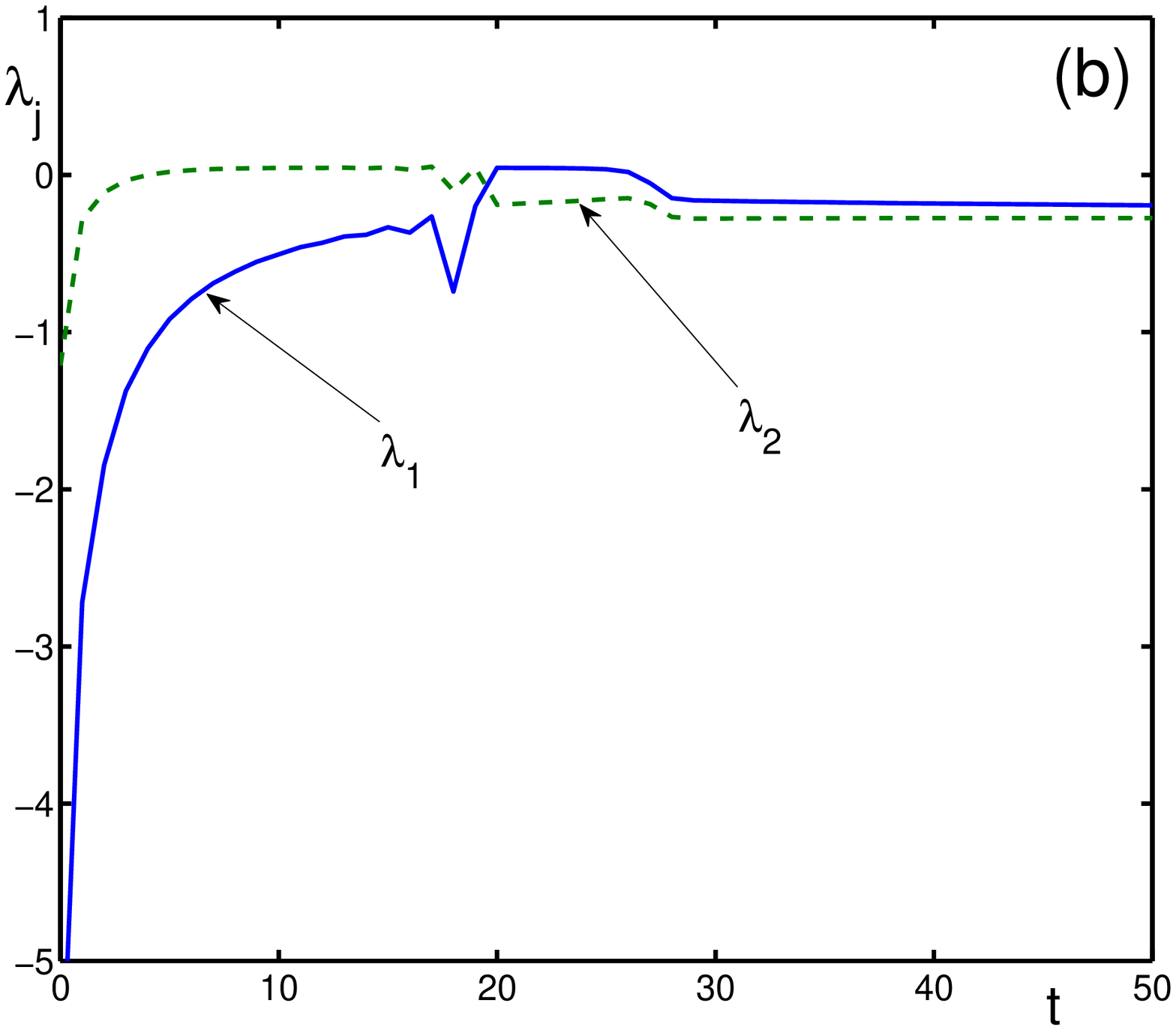} } }
\caption{Reconstructive memory under conflict. Dynamics of the probabilities (a) and
local Lyapunov exponents (b) for the initial conditions $f_1 = 0.2 < f_2 = 0.55$, with
$q_1(0) = 0.3$ and $q_2(0) = -0.2$, so that $p_1(0) = 0.5 > p_2(0) = 0.35$.
}
\label{fig:Fig.10}
\end{figure*}

\begin{figure*}[!t]
\vspace{9pt}
\centerline{
\hbox{ \includegraphics[width=7.5cm]{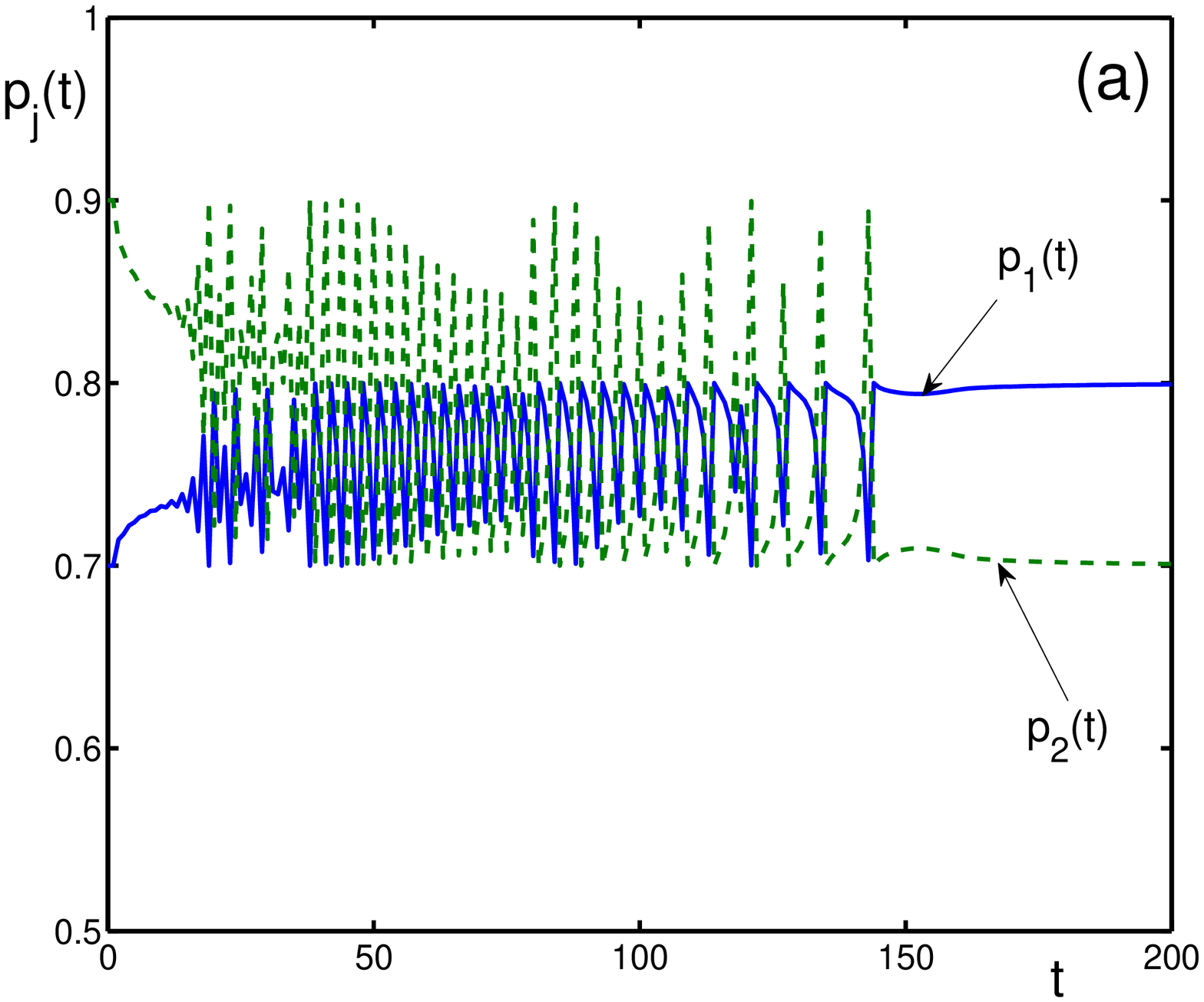}
\hspace{1.5cm}
\includegraphics[width=7.5cm]{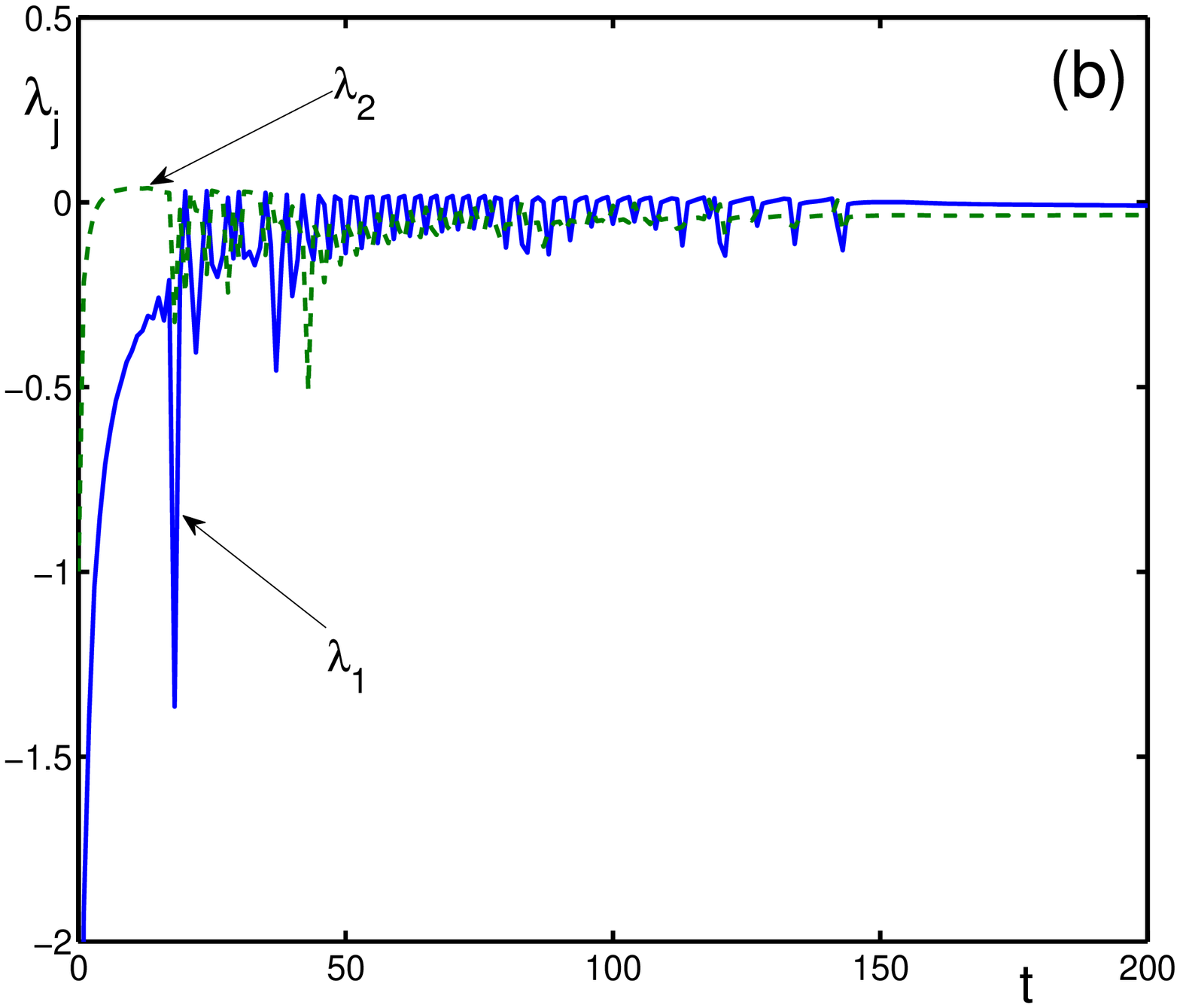} } }
\caption{Reconstructive memory under conflict. Dynamics of the probabilities (a) and
local Lyapunov exponents (b) for the initial conditions $f_1 = 0.8 > f_2 = 0.7$, with
$q_1(0) = -0.1$ and $q_2(0)=0.2$, so that $p_1(0)=0.7<p_2(0)=0.9$. At the intermediate
stage, the probabilities exhibit chaotic oscillations.
}
\label{fig:Fig.11}
\end{figure*}

\section{Agents with Short-Term Memory}

The other limiting case is the society of agents with very short-term memory,
remembering only the information from the last temporal step and keeping no track
of any previous information gains. This is described by the local memory operator
$\hat{\varphi}_j(t,k) = \delta_{kt}$. As a result, the total information coincides with
the information gain from the last step, 
\be
\label{46}
M_j(t) = \mu_j(t) \; .
\ee
All other equations are the same as above, with the same initial conditions. In 
particular, at the initial time there is no yet any information, as has been assumed 
above, so that in the present case we have
\be
\label{47}
M_j(0) = \mu_j(0) = 0  \; .
\ee

Numerical analysis shows that there exist three types of dynamics. One is 
a smooth tendency to limiting states from below or from above, as is illustrated 
in Fig. 12. The second type is the tendency to limiting states through several or 
a number of oscillations, as in Fig. 13. And the third kind of dynamics is the 
occurrence of everlasting oscillations, intersecting or not intersecting with 
each other, as is shown in Fig. 14. The existence of these three types of dynamics
happens for positive as well as for negative attraction factors. The limits of the
trajectories at infinite time, when they exist, depend on initial conditions and do 
not equal the related utility factors. Slightly varying the initial conditions leads to 
small finite changes in the limiting trajectory values, so that the motion is Lyapunov 
stable. But it is not asymptotically stable, contrary to the cases of societies of 
agents with long-term or reconstructive memories. The existence of stable 
everlasting oscillations is also typical only for the agents with short-term memory, 
but does not occur for agents with other types of memory.

\begin{figure*}[!t]
\vspace{9pt}
\centerline{
\hbox{ \includegraphics[width=7.5cm]{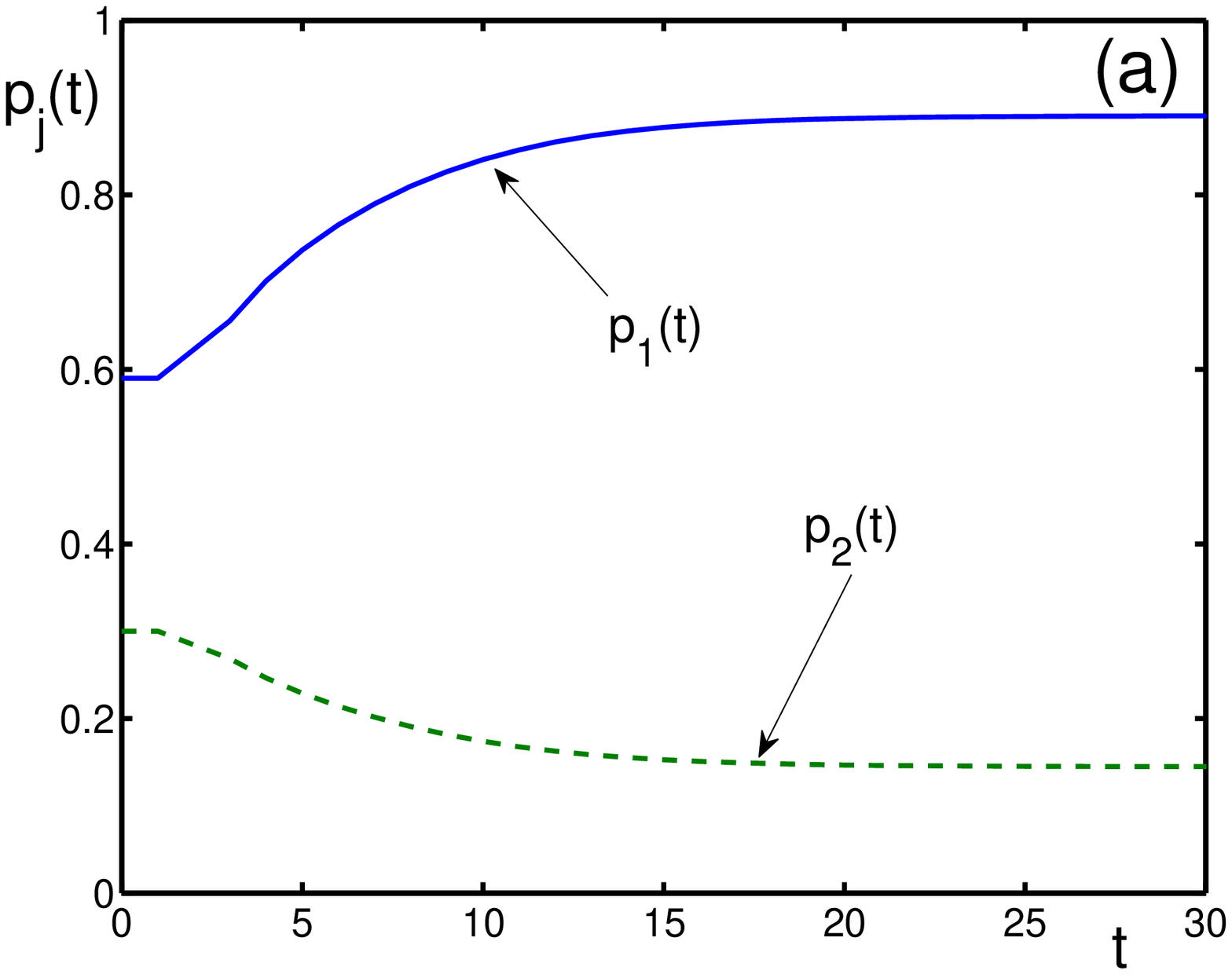}
\hspace{1.5cm}
\includegraphics[width=7.5cm]{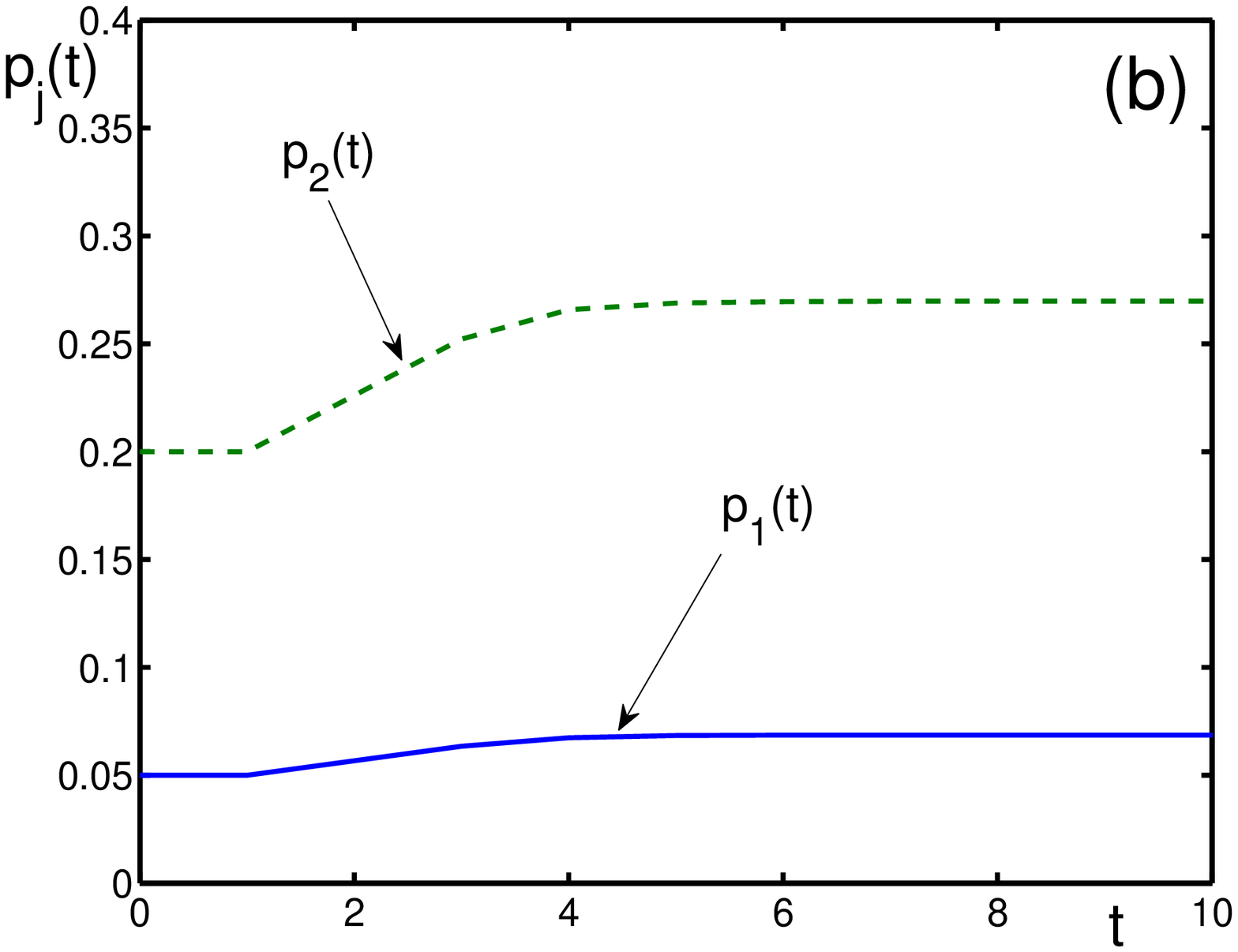} } }
\caption{Short-term memory. Dynamics of the probabilities for the initial 
conditions: (a) $f_1=0.99$, $f_2=0.1$, with $q_1(0)=-0.4$ and $q_2(0)=0.2$, so 
that $p_1(0)=0.59>p_2(0)=0.3$, and (b) $f_1=0.2$, $f_2=0.6$, with $q_1(0)=-0.15$ 
and $q_2(0)=-0.4$, so that $p_1(0)=0.05<p_2(0)=0.2$. The probabilities exhibit 
monotonic behavior irrespectively of conflict or no-conflict initial conditions.
}
\label{fig:Fig.12}
\end{figure*}

\begin{figure*}[!t]
\vspace{9pt}
\centerline{
\hbox{ \includegraphics[width=7.5cm]{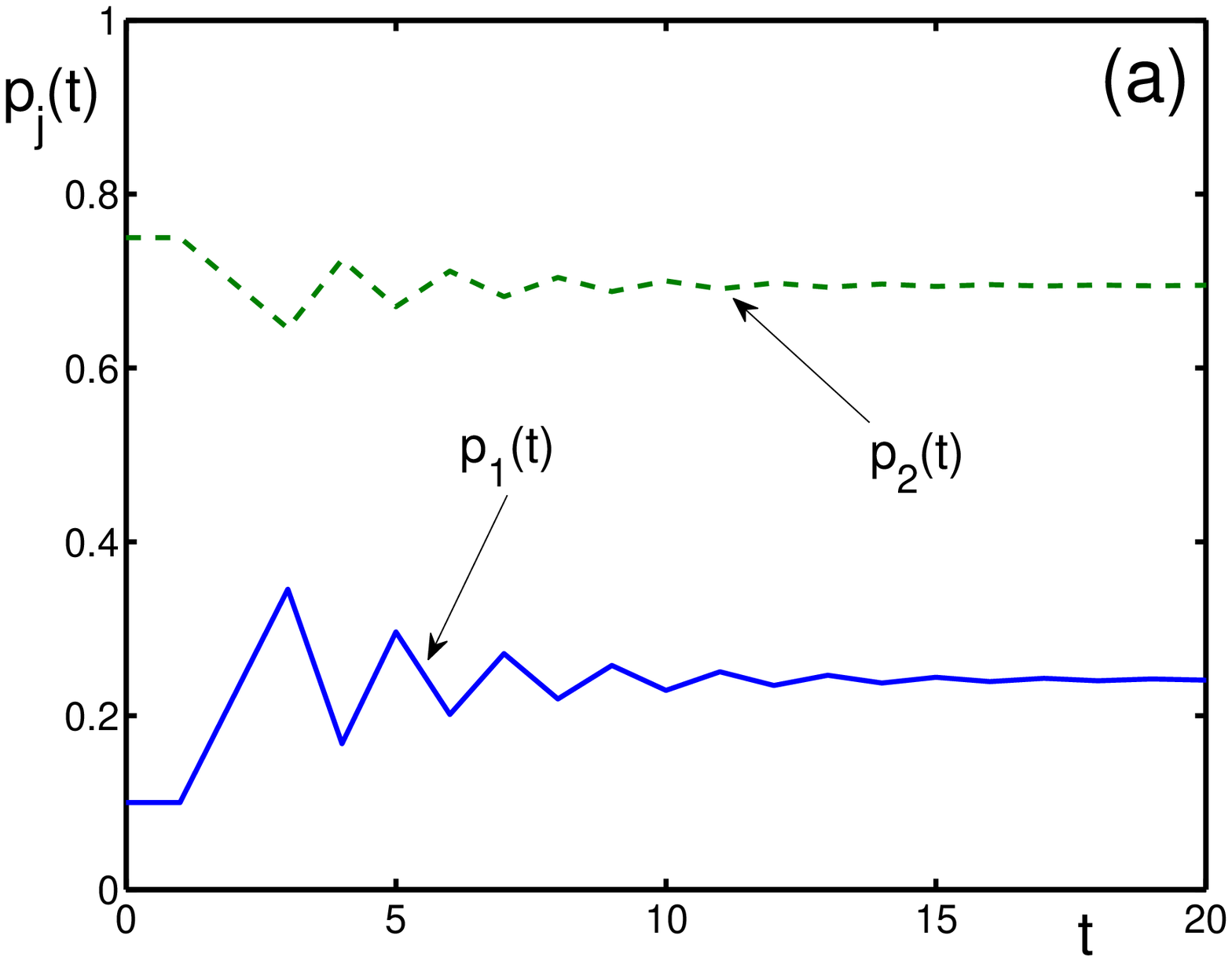}
\hspace{1.5cm}
\includegraphics[width=7.5cm]{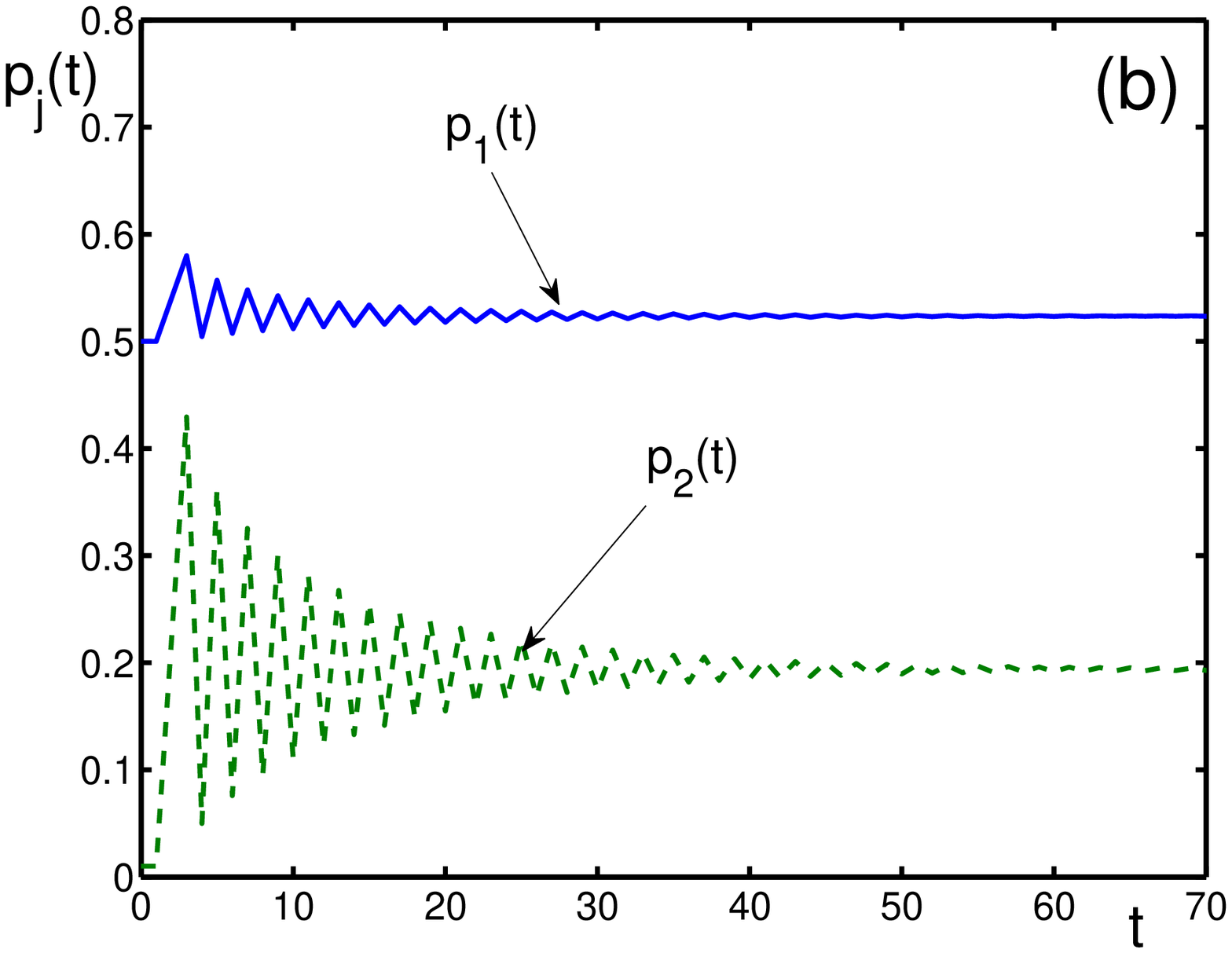} } }
\caption{Short-term memory. Dynamics of the probabilities for the initial 
conditions (a) $f_1=0.5$, $f_2=0.6$, with $q_1(0)=-0.4$ and $q_2(0)=0.15$, so 
that $p_1(0)=0.1<p_2(0)=0.75$, and (b) $f_1=0.6$, $f_2=0.9$, with $q_1(0)=-0.1$ 
and $q_2(0)=-0.89$, so that $p_1(0)=0.5>p_2(0)=0.01$. The probabilities exhibit 
decaying oscillations irrespectively of conflict or no-conflict initial conditions.
}
\label{fig:Fig.13}
\end{figure*}

\begin{figure*}[!t]
\vspace{9pt}
\centerline{
\hbox{ \includegraphics[width=7.5cm]{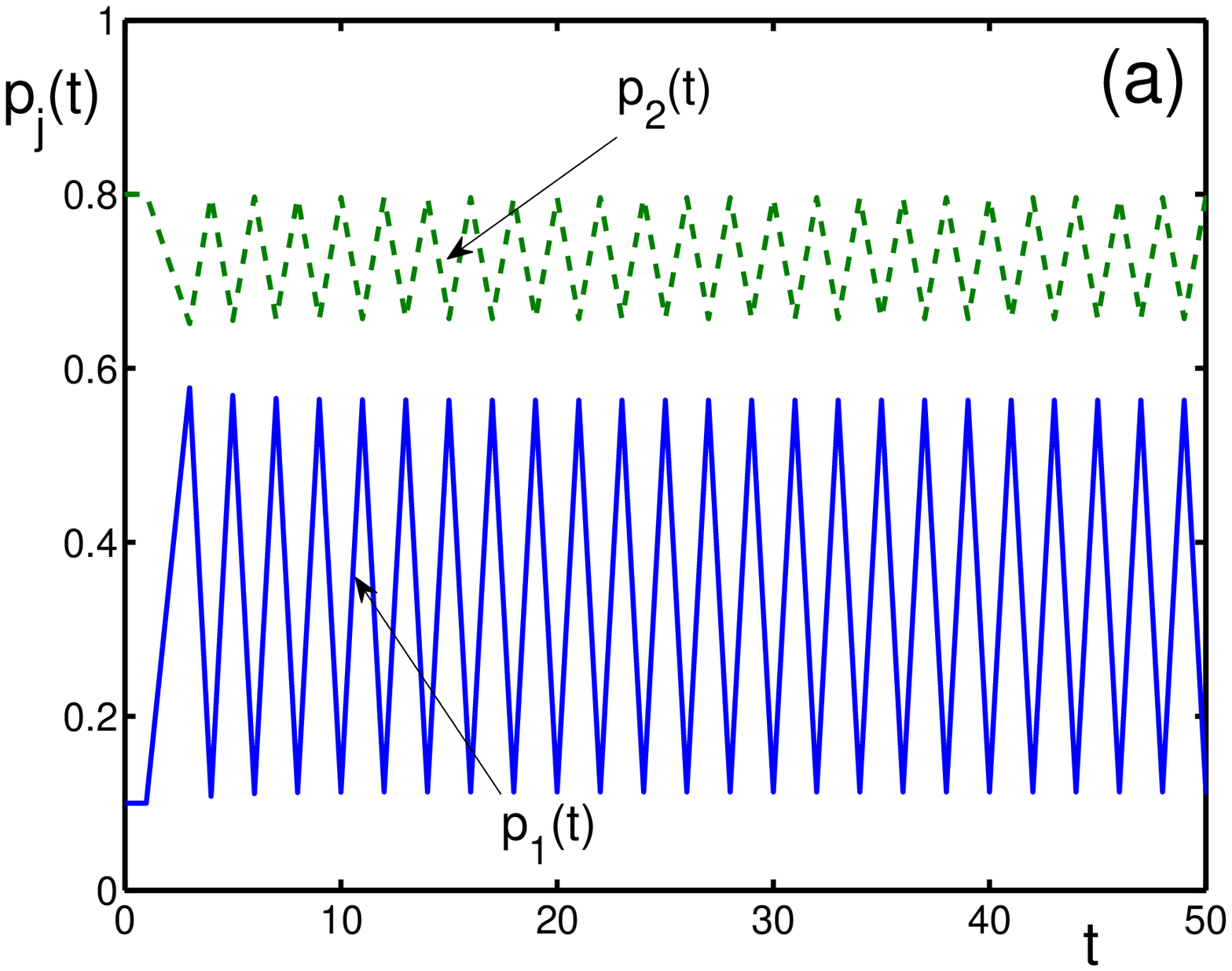}
\hspace{1.5cm}
\includegraphics[width=7.5cm]{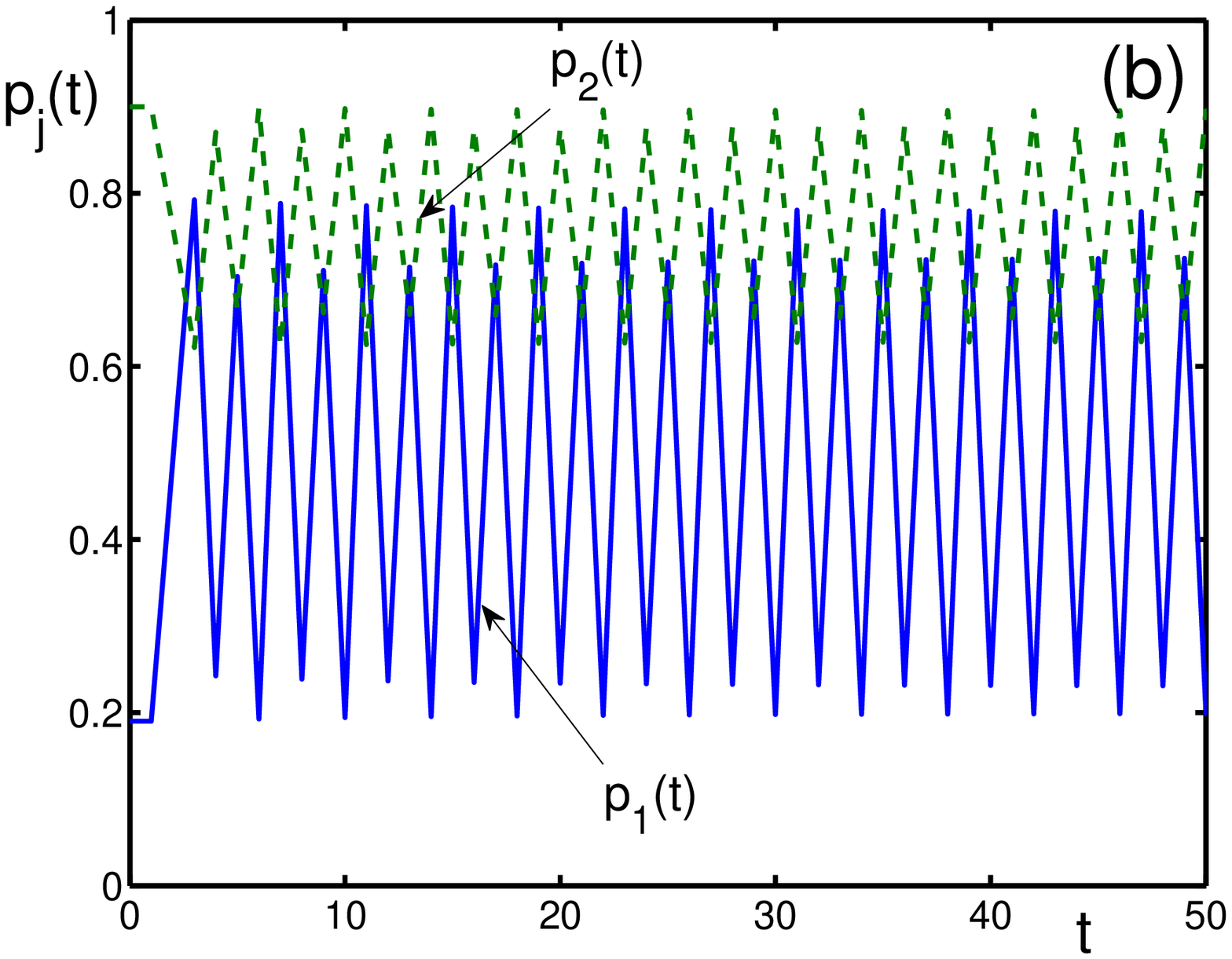} } }
\caption{Short-term memory. Dynamics of the probabilities for the initial 
conditions: (a) $f_1 = 0.8$, $f_2 = 0.6$, with $q_1(0) = -0.7$ and $q_2(0)=0.2$, 
and (b) $f_1 = 0.99$, $f_2 = 0.5$, with $q_1(0) = -0.8$ and $q_2(0)=0.4$. 
 The probabilities exhibit everlasting oscillations without intersection (a) or with 
intersecting probabilities (b). The everlasting oscillations exist irrespectively of 
conflict or no-conflict initial conditions.
}
\label{fig:Fig.14}
\end{figure*}

\section{Societies of Decision Makers as Intelligence Networks}

A society of agents generating, in the course of decision making, probability 
distributions over a given set of prospects, and interacting with each other 
through information exchange, is analogous to a network, which can be termed 
a {\it social information network}. Recall that the mathematics of taking 
decisions according to the rules of QDT is equivalent to the activity of 
an artificial intelligence, since an artificial intelligence, mimicking human 
cognition, has to take account of the conscious-subconscious duality typical 
of human brains \cite{YS_1,YS_119,YS_2}. And this duality is well represented 
by QDT, where the prospect probabilities consist of two terms, utility factor and 
attraction factor.      

Therefore the society of decision makers, acting in the frame of QDT, is 
equivalent to a {\it quantum information network} or {\it quantum intelligence network}. 
In other words, such a network can characterize a composite artificial intelligence, 
or {\it superintelligence}, formed by an assembly of intelligences. In that sense, 
it is essentially more complicated than other known networks.        

There exists a large variety of networks: internet and web networks, computer 
networks, social networks, business networks, radio and television networks, 
electric networks, phone-call networks, citation networks, linguistic networks, 
ecological and biological networks, cellular networks, protein networks, neural 
networks, etc. \cite{Wasserman_55,Bishop_56,Gurney_57,Scott_58,Albert_59}.
All above mentioned networks are characterized by classical models. There are 
also quantum networks that are usually modelled by quantum spin systems or 
exciton systems that can be reduced to spin models \cite{Nikolopoulos_60,Blok_61}.

In mathematical terms, a network is defined as follows. There is a set 
$\mathbb{A} = \{a_j\}$ of nodes, or vertices, or agents, enumerated by an index 
$j = 1,2, \ldots, N$ and a set $\mathbb{L} =\{ l_{ij} \}$ of edges, or lines, or 
links, or arcs, connecting the nodes. The pair $\{\mathbb{A}, \mathbb{L}\}$ is 
called a graph. The graph is termed `directed' if there is a map
$m : \mathbb{L} \longrightarrow \mathbb{R}_+ \equiv [0, \infty)$. For a directed 
graph, an edge $l_{ij}$ may be different from $l_{ji}$. A pair of edges $l_{ij}$ 
and $l_{ji}$, connecting two nodes, is called a circuit. Generally, a network is 
the triple $\{\mathbb{A}, \mathbb{L}, m\}$ representing a directed graph. In 
standard quantum networks, nodes are represented by quantum operators, usually
by spin or quasi-spin operators, and their interactions by parameters of an 
interaction matrix. Signals are characterized by wave functions. A node operator 
transforms a given wave function into another function, thus realizing a gate 
\cite{Nikolopoulos_60,Blok_61}.

A society of decision makers, or an assembly of artificial intelligences,
functioning by the rules of QDT, can be classified as a network. The set of nodes 
is represented by decision makers, whose links are described by the exchanges 
of information. The corresponding graph is directed, since the information received 
by an agent, generally speaking, can differ from that received by other agents. 
Keeping in mind that the network dynamics generates probability distributions varying 
with time, it is straightforward to interpret this activity as time-dependent decision 
making \cite{Nicolis_62}. The delayed information exchange and link directionality 
reflect the causality of interactions \cite{Granger_63}. The network dynamics 
describes the information flow \cite{Shaw_64}. This is why the intelligence networks 
can also be termed the information networks. 

The principal difference of a {\it quantum intelligence network} from the usual 
quantum networks, with the nodes given by spins or atoms, is that each agent in an 
intelligence network makes decisions, while the standard nodes, like spins, do not 
do this. The type of intelligence networks we propose represents networks 
of agents acting by the quantum rules of QDT, with outcomes that are not just simple 
two-bit signals, as yes or no, and which could be modelled by spin systems with spins 
up or down. In contrast, in the quantum network proposed here, each agent generates
a quantum probability measure over a given set of prospects. Being based on common 
mathematics, the intelligence networks can characterize either an assembly of 
interacting human decision makers, or a cluster of several quantum computers, or the 
activity of a composite artificial quantum intelligence consisting of several parts, each 
of which is an intelligence itself. The network dynamics models the formation of a 
collective decision as a collective outcome developing in the process of information 
exchange.

\section{Example of Dynamic Decision Making}
 
\subsection{Position of the problem}

To illustrate the usage of the theory, we treat below a particular case of dynamic 
decision making. There are three main problems in describing such a repeated 
decision making. 

\begin{itemize}

\item[(i)] At the beginning, decisions under uncertainty often contradict utility theory, 
as is discussed in the Introduction. Decision makers often choose the prospect with 
smaller utility factor, in contradiction with the prescription of utility theory. Why this 
is so and how to correctly predict the behavioral choice of decision makers at the 
initial stage? 

\item[(ii)] There exist numerous empirical works \cite{Charness_66,Blinder_67,Cooper_68,
Charness_69,Charness_70,Chen_71,Charness_72,Kuhberger_73,
Tsiporkova_117,Liu_118} showing that the difference between the probability of 
choosing a prospect and the utility factor decrease with time. For a society of 
decision makers, we define the experimental probability of a given prospect as the 
fraction of agents choosing that prospect. How to prove mathematically that this 
difference between the prospect probability and utility factor does diminish with time?

\item[(iii)] Empirical works \cite{Charness_66,Blinder_67,Cooper_68,Charness_69,
Charness_70,Chen_71,Charness_72,Kuhberger_73,Tsiporkova_117,Liu_118} 
also show that, even if at the initial time the behavioral probabilities strongly deviate 
from the prescription of utility theory, in the long run they converge to the related utility 
factors. The convergence of initially dispersed opinions to a common consensus is 
the basis of the so-called Delphi method that has been devised in order to obtain the 
most reliable opinion consensus of a group of experts by subjecting them to a series 
of questionnaires interspersed with discussions providing feedback
\cite{Dalkey_119,Kozak_120}. The empirical studies raise the following question: How 
to describe the fact that, for real decision makers exchanging information, the initially 
dispersed opinions converge to a common consensus and, furthermore, the limiting 
prospect probabilities converge to an objective utility factor?  

\end{itemize}        

We illustrate the above questions by a concrete example involving the dynamic 
disjunction effect, and compare our theoretical predictions with empirical data. Since 
the disjunction effect is specified as a violation of the {\it sure-thing principle} 
\cite{Savage_121}, we first briefly recall the meaning of this principle. Then we give 
a real-life example of the effect, concentrating on the case of a composite game 
studied by Tversky and Shafir \cite{Tversky_122}. After this, we analyze the disjunction 
effect dynamics by using the Tversky and Shafir data to specify the numerical values 
needed in the numerical solution of our equations to obtain the limiting values of the 
corresponding prospect probabilities.   

\subsection{Sure-thing principle}

Let us consider a two-stage composite prospect. In the first stage, the events 
$B_1$ and $B_2$ can occur. One can either know for sure that one of these concrete 
events has occurred or one can only be aware that one of them has happened, not 
knowing which of them actually occurred. The latter case can be denoted as 
$B = \{B_1,B_2\}$. At the second stage, either an event $A_1$ or $A_2$ occurs. 
We are thus confronted with the composite prospects $\pi_n = A_n \bigotimes B$, 
with $n = 1,2$. The sure-thing principle \cite{Savage_121} states: {\it If the alternative $A_1$ 
is preferred to the alternative $A_2$, when an event $B_1$ occurs, and it is also 
preferred to $A_2$, when an event $B_2$ occurs, then $A_1$ should be preferred 
to $A_2$, when it is not known which of the events, either $B_1$ or $B_2$, has occurred}. 

This principle is easily illustrated in classical probability theory, where the probability
of the prospect $\pi_n$ is
\be
\label{48}
 f(\pi_n) = p(A_nB_1) + p(A_nB_2) \;  .
\ee
Then, if $p(A_1B_\alpha) > p(A_2B_\alpha)$ for $\alpha = 1,2$, it follows that
$f(\pi_1) > f(\pi_2)$, which explains the sure-thing principle.

\subsection{Disjunction effect}

However, empirical studies have discovered numerous violations of the sure-thing
principle, which was called disjunction effect 
\cite{Tversky_122,Croson_123,Lambdin_124,Li_125}. Such violations are typical for 
two-step composite games of the following structure. First, a group of agents takes
part in a game, where each agent can either win (event $B_1$) or loose (event $B_2$),
with equal probability $p(B_\alpha) = 0.5$. They are then invited to participate in a 
second game, having the right either to accept the second game (event $A_1$) or to 
refuse it (event $A_2$). The second stage is realized in different variants: One 
can either accept or decline the second game under the condition of knowing the result 
of the first game. Or one can either accept or decline the second game without knowing 
the result of the first game. We define the probabilities, as usual, in the frequentist 
sense as the fractions of individuals taking the corresponding decision. 

In their experimental studies, Tversky and Shafir \cite{Tversky_122} find that the 
fraction of people accepting the second game, under the condition that the first was 
won, is $p(A_1|B_1) = 0.69$ and the fraction of those accepting the second game, 
under the condition that the first was lost is $p(A_1|B_2) = 0.59$. From the definition 
of conditional probabilities, one has the normalization   
$$
 p(A_1|B_\al) + p(A_2|B_\al) = 1 \qquad (\al=1,2) \; ,
$$
which yields $p(A_2|B_1) = 0.31$ and $p(A_2|B_1) = 0.41$. Therefore the related 
joint probabilities $p(A_nB_\alpha) = p(A_n|B_\alpha) p(B_\alpha)$ are 
$$
p(A_1B_1) = 0.345 \; , \qquad p(A_1B_2) = 0.295 \; ,
$$
$$
 p(A_2B_1) = 0.155 \; , \qquad p(A_2B_2) = 0.205 \; .
$$
According to equation (\ref{48}), one gets 
$$
f(\pi_1) = 0.64 \; , \qquad f(\pi_2) = 0.36 \; .
$$
This implies that, by classical theory, the probability of accepting the second game,
not knowing the results of the first one, is larger than that of refusing the second 
game, not knowing the result of the first game.    

Surprisingly, in their experiments, Tversky and Shafir \cite{Tversky_122} observe that 
human decision makers behave opposite to the prescription of the sure-thing principle, 
with the majority refusing the second game, if the result of the first one is not known,
$$
 p_{exp}(\pi_1) = 0.36 \; , \qquad p_{exp}(\pi_2) = 0.64 \; .
$$
 
But in QDT, such a paradox does not appear. Recall that, in QDT, the probability of a
prospect is the sum of $f(\pi_n)$ and $q(\pi_n)$. The sign of the attraction factor is
prescribed by the uncertainty aversion \cite{YS_1,YS_119,YS_2,YS_17}, while its 
noninformative prior value is found \cite{YS_1,YS_119,YS_2} to satisfy the quarter law,
with the average absolute value of the attraction factor $|q| = 0.25$. Thus, in QDT, 
we have
$$
p(\pi_1) = f(\pi_1) - 0.25 = 0.39 \; , 
$$
$$
 p(\pi_2) = f(\pi_2) + 0.25 = 0.61 \; .
$$
These theoretical results for $p(\pi_n)$ are in very good agreement with the experimental 
data $p_{exp}(\pi_n)$ by Tversky and Shafir, actually being indistinguishable within the
accuracy of experimental statistics.  

Similar results, demonstrating the disjunction effect, have been obtained in a variety of 
other empirical studies having the same two-stage games structure
\cite{Tversky_122,Croson_123,Lambdin_124,Li_125}. All such paradoxes find a
simple explanation in QDT, similarly to the case treated above.

\subsection{Disjunction-effect dynamics}

We now take the data from the previous subsection as initial conditions of our 
dynamical equations of Sec. 5 and study the time dependence of the decision 
makers' opinions.

There are two prospects. One is the prospect $\pi_1 = A_1 \bigotimes B$ of accepting 
the second game not knowing the result of the first game. And the second is the prospect 
$\pi_2 = A_2 \bigotimes B$ of refusing from the second game, when the result of the first 
game is not known. As is explained in Sec. 5, using the normalization conditions 
for the probabilities, it is sufficient to consider one of the prospects, say $\pi_1$. 
 
Taking into account the heterogeneity of agents, we assume that there are two main 
groups of agents differing in their initial opinions. The corresponding prospect probabilities
are denoted as $p_j(t) \equiv p_j(\pi_1,t)$. Respectively, we use the abbreviated notation
for the utility factors $f_j(t) \equiv f_j(\pi_1,t)$ and for the attraction factors  
$q_j(t) \equiv q_j(\pi_1,t)$. The utility factor, being an objective quantity, is assumed to be
constant and defined as in the previous subsection: $f_j(t) = 0.64$.  
As a result of the inequalities $0 \leq p_j(t) \leq 1$ for the probabilities, the attraction factor 
can vary in the range
$$
 - f_j(t) \leq q_j(t) \leq 1 - f_j(t) \;  ,
$$
which in the present case implies that $- 0.64 < q_j(t) < 0.36$. The initial values of 
the attraction factors must fall within this interval. 

We consider the more realistic situation where the decision makers take repeated 
decisions  and remember their previous choices. We solve the 
evolution equations of Sec. 5 for this type of memory, taking as initial conditions for 
the attraction factors uniformly distributed values taken in the interval $(- 0.64, 0.36)$. 
The typical obtained behavior of the prospect probabilities for different initial conditions 
are shown in Fig. 15. 

Two important conclusions follow from the temporal behavior of the solutions. First, 
for any initial conditions, the difference between the probabilities $p_j(t)$ and the utility
factor decreases with time. Second, all solutions, for arbitrary initial conditions, tend to 
the same limit $0.64$. This demonstrate that, despite a large variation in the initial 
opinions, the agents come to a common consensus by exchanging information in a 
multi-step procedure.

\begin{figure*}[!t]
\vspace{9pt}
\centerline{
\hbox{ \includegraphics[width=8.5cm]{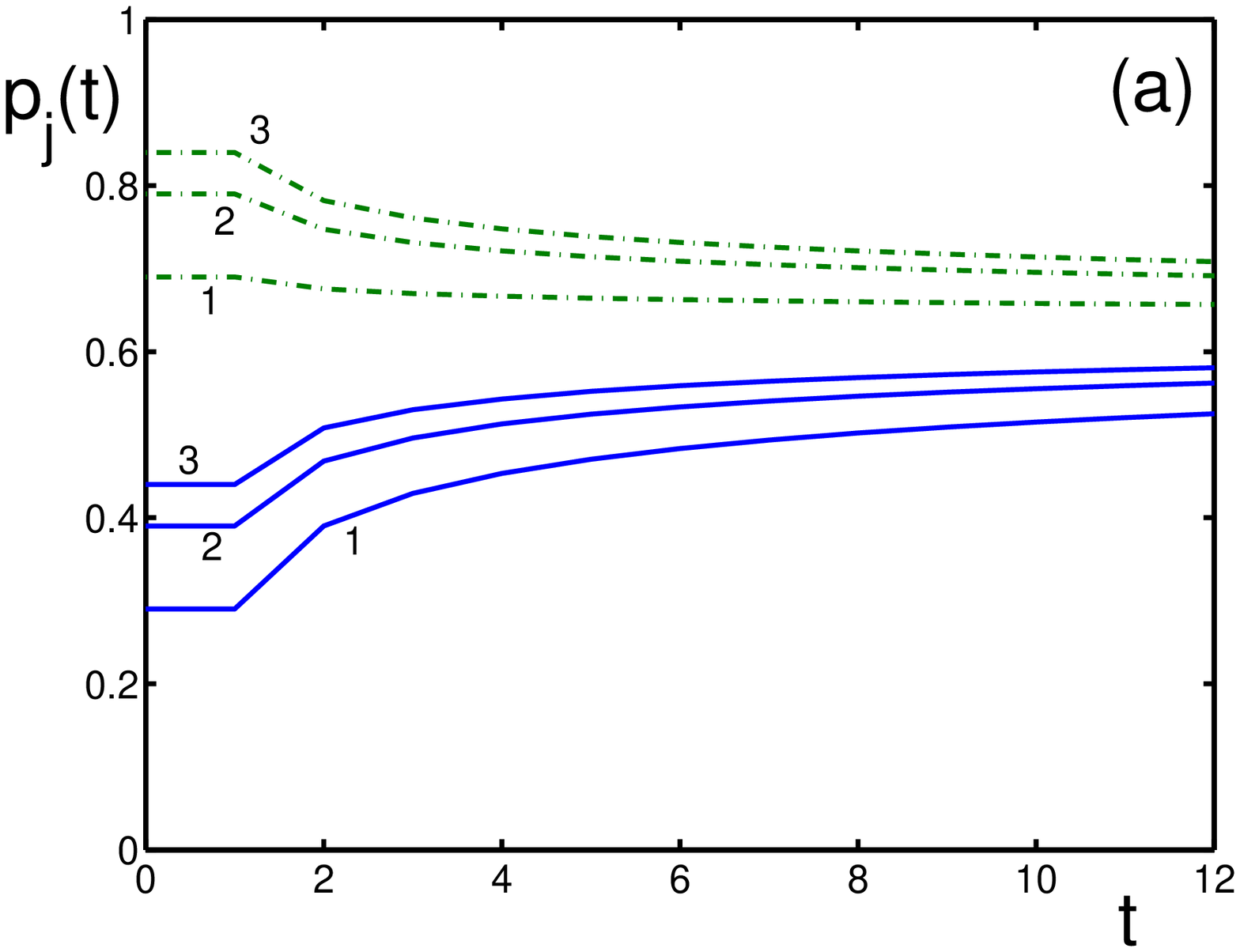}
\hspace{0.5cm}
\includegraphics[width=8.5cm]{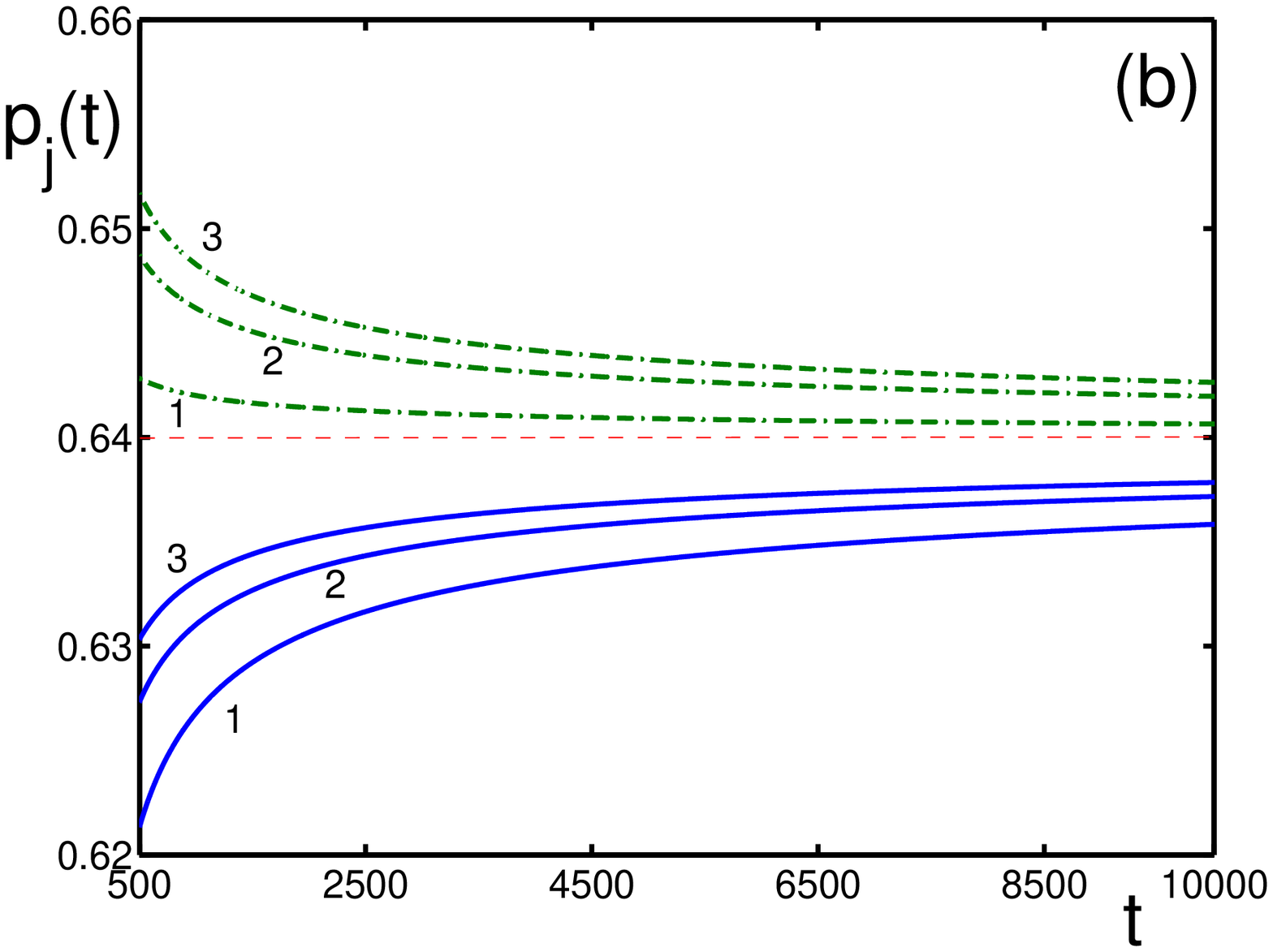} } }
\caption{ Dynamic disjunction effect. The prospect probabilities $p_1(t)$ (solid lines) 
and $p_2(t)$ (dashed-doted lines) as functions of time $t$ for different initial attraction 
factors: (1) $q_1(0)=-0.35$ and $q_2(0)=0.05$; (2) $q_1(0)=- 0.25$ and $q_2(0) = 0.15$;
(3) $q_1(0) = - 0.2$ and $q_2(0) = 0.2$. Figures (a) and (b) show the behaviour of the
probabilities in the initial and long-term time intervals $t\in [0,12]$ and $t\in [500,10000]$,
respectively. When $t \ra \infty$, then $p_1(t) \ra 0.64$ from below, and $p_2(t) \ra 0.64$
from above.
}
\label{fig:Fig.15}
\end{figure*}

\section{Conclusion}

We have suggested a model of a society of decision makers taking their decisions
according to the quantum decision theory. This model generalizes the theory, formulated 
earlier for one-step decisions of single decision makers to multistep decision making 
of interacting agents in a society. Such a society of decision makers, acting according 
to the rules of QDT, represents a new kind of networks whose nodes are the QDT agents, 
interacting through the exchange of information. Each agent generates a probability 
measure over a set of prospects. The generated probabilities are defined according to 
quantum rules, which results in their representation as sums of two terms, positive 
definite and sign indefinite. The quantum-classical correspondence principle makes it 
possible to interpret the positive-definite term as a classical probability or utility factor. 
The sign-indefinite term, having purely quantum origin, represents the attractiveness 
of the given prospects, and is referred to as the attraction factor. The utility factor can 
be considered as given and time-invariant. In contrast, the attraction factor is random 
at the initial time and varies with time due to the information exchange between the agents.

It is worth stressing that our approach is principally different form stochastic decision 
theory. The variants of such theories are usually based on deterministic decision theories
complemented by random variables with given distributions \cite{Cockett_114,Blavatskyy_44}. 
Therefore such stochastic theories inherit the same problems as deterministic theories 
embedded into them. Moreover, stochastic theories are descriptive, containing fitting 
parameters that need to be defined from empirical data. In addition, different stochastic 
specifications of the same deterministic core theory may generate very different, and 
sometimes contradictory, conclusions \cite{Loomes_115}.  In stochastic decision theory,
random terms are usually added to expected utility, while in our case the quantum prospect  
probability acquires an additional term, named attraction factor. What is the most important,
our approach is not descriptive and does not contain fitting parameters. 

Our model of a society of agents interacting through information exchange is principally 
new and has never been considered before to the best of our knowledge. We have 
suggested the first model describing the opinion dynamics of real decision makers 
subject to behavioral effects. 

The probability dynamics depends on the form of the total information accumulated 
by each agent. We have analyzed three qualitatively different limiting cases, long-term 
memory, reconstructive memory filling the past gaps on the basis of the most recent
information gain, and short-term memory. In the case of long-term memory, the 
probability dynamics is smooth. When the initial conditions are not conflicting, the 
probabilities tend to their related utility factors $(p_j \ra f_j)$. But when the initial 
conditions are conflicting, the probabilities tend to a consensual common limit 
$(p_j \ra p^*)$. In all the cases, the motion is asymptotically Laypunov stable. 

The probabilities for the agents with reconstructive memory always tend to their 
related utility factors $(p_j \ra f_j)$, never exhibiting consensus. At intermediate stages 
of their dynamics, the probabilities can experience oscillations, with positive local 
Lyapunov exponents, which implies local instability. But in the long run, the dynamics 
becomes asymptotically Lyapunov stable.

The society of agents with short-term memory behaves rather differently from the 
behavior of the agents in the previous cases. Two types of dynamics can occur. The 
probability trajectories can tend to fixed points $(p_j \ra p_j^*)$ strongly depending on 
initial conditions, so that the motion is not asymptotically Lyapunov stable, although 
Lyapunov stable. These fixed points are different from the utility factors. The other 
type of motion is characterized by everlasting oscillations, which seems to be natural 
for agents with short-term memory, where there is no information accumulation, because 
of which the agents cannot make precise stable decisions. Also, for such agents, a 
consensus is impossible. 

The dynamics of decisions is due to the time dependence of the attraction factors,
since the utility factors have been taken as invariants representing the objective 
utility of the related prospects. The decrease of the attraction factors with time, 
due to the exchange of information between decision makers, amounts to a reduction 
of the inconsistencies with utility theory. This decay of the deviations from utility theory, 
caused by the information exchange between decision makers, has been observed 
in many empirical studies \cite{Charness_66,Blinder_67,Cooper_68,Charness_69,
Charness_70,Chen_71,Charness_72,Kuhberger_73,Dalkey_119,Kozak_120}.

From another side, as stressed above, a society of decision makers, acting in
the frame of QDT, is equivalent to an intelligence network, where the agents take 
decisions respecting the conscious-subconscious duality. Since this duality is typical 
of human decision makers, the networks, attempting to mimic the activity of human 
brains, are to be based on the rules of QDT. This should be taken into account in
creating artificial intelligence and networks of artificial intelligences. 

The usage of the approach is illustrated by a concrete realistic example involving 
the dynamic disjunction effect. Treating this effect in the frame of our theory yields 
three important predictions. (i) The initial values of the prospect probabilities are 
predicted in very good agreement with empirical data. (ii) The difference between 
the initial probabilities and the corresponding utility factors monotonically decrease 
with time. (iii)  In a society of heterogeneous agents with long-term memory and 
randomly chosen initial conditions, all prospect probabilities tend to a common limit 
coinciding with the utility factor, thus demonstrating the existence of a consensus 
as a result of the repeated information exchange. These predictions are in agreement 
with empirical data.

\section*{Acknowledgement}

We acknowledge financial support from the ETH Z\"{u}rich Risk Center.

\end{document}